\documentclass[a4paper,12pt]{article}
\usepackage{amsmath}
\usepackage{amssymb}
\usepackage{fullpage}
\usepackage{verbatim}

\usepackage[T1,OT1]{fontenc}

\usepackage{graphicx}

\allowdisplaybreaks[3]

\usepackage[hang]{footmisc}
\usepackage[compress,square,numbers]{natbib}
\usepackage[colorlinks,linktocpage,linkcolor=blue,citecolor=blue,urlcolor=blue]{hyperref}

\newcommand{\e}{\mathrm{e}}

\newcommand{\nl}{\notag \\ &\quad\,}

\begin{document}

\numberwithin{equation}{section}

\thispagestyle{empty}

\begin{flushright}
\small LMU-ASC 65/16\\
\small ITP-UH-25/16\\
\normalsize
\end{flushright}
\vspace*{1cm}

\begin{center}

{\LARGE \bf A Universal Tachyon in Nearly No-scale}

\vspace{0.5cm}

{\LARGE \bf de Sitter Compactifications}

\vspace{2cm}
{\large Daniel Junghans${}^{1}$ and Marco Zagermann${}^{2,3}$}\\

\vspace{1cm}
${}^1$ Arnold-Sommerfeld-Center f{\"{u}}r Theoretische Physik\\
Department f{\"{u}}r Physik, Ludwig-Maximilians-Universit{\"{a}}t M{\"{u}}nchen\\
Theresienstra\ss e 37, 80333 M{\"{u}}nchen, Germany\\

\vspace{0.5cm}
${}^2$ Institut f{\"u}r Theoretische Physik \&\\
${}^3$ Riemann Center for Geometry and Physics\\
Leibniz Universit{\"a}t Hannover, Appelstra{\ss}e 2, 30167 Hannover, Germany\\

\vspace{1cm}
{\upshape\ttfamily daniel.junghans@lmu.de, marco.zagermann@itp.uni-hannover.de}\\

\vspace{1.5cm}

\begin{abstract}
\noindent We investigate de Sitter solutions of $\mathcal{N}=1$ supergravity with an F-term scalar potential near a no-scale Minkowski point, as they may in particular arise from flux compactifications in string theory. We show that a large class of such solutions has a universal tachyon with $\eta \le -\frac{4}{3}$ at positive vacuum energies, thus forbidding meta-stable de Sitter vacua and slow-roll inflation. The tachyon aligns with the sgoldstino in the Minkowski limit, whereas the sgoldstino itself is generically stable in the de Sitter vacuum due to mass mixing effects. We specify necessary conditions for the superpotential and the K{\"{a}}hler potential to avoid the instability.
Our result may also help to explain why the program of classical de Sitter hunting has remained unsuccessful, while constructions involving instantons or non-geometric fluxes have led to various examples of meta-stable de Sitter vacua.
\end{abstract}

\end{center}

\newpage

\section{Introduction}

Attempts to construct explicit meta-stable de Sitter (dS) vacua in string theory face a number of well-known theoretical and computational challenges. In recent years, various approaches have addressed these challenges and led to a number of interesting constructions and successful models (see, e.g., \cite{Kachru:2003aw, Burgess:2003ic, Balasubramanian:2005zx,Rummel:2011cd, Louis:2012nb,Cicoli:2012fh, Cicoli:2013cha, Blaback:2013qza, Danielsson:2013rza, Rummel:2014raa, Braun:2015pza, Kallosh:2014oja, Marsh:2014nla, Guarino:2015gos, Retolaza:2015nvh, Dong:2010pm, Dodelson:2013iba, deCarlos:2009fq, deCarlos:2009qm, Dibitetto:2010rg, Danielsson:2012by, Blaback:2013ht, Damian:2013dq, Damian:2013dwa, Hassler:2014mla, Blaback:2015zra, Cicoli:2013rwa, Parameswaran:2006jh, Kounnas:2014gda, Achucarro:2015kja, Cicoli:2015ylx, Ben-Dayan:2013fva, Blumenhagen:2015kja, Blumenhagen:2015xpa}). Despite this encouraging progress, however, it remains fair to say that we are still far away from a full classification of the whole landscape of possible dS vacua in string theory. Many interesting questions are related to this issue: Which other mechanisms for dS vacua, if any, exist in string theory? What are the minimal ingredients a compactification requires in order to admit dS vacua? Is it possible to construct simpler models than those already known, perhaps even some that are fully explicit at the 10d level? How are observables such as the cosmological constant, moduli masses or the supersymmetry breaking scale distributed across the landscape (see, e.g., \cite{Sumitomo:2012wa, Sumitomo:2012vx, Tye:2016jzi} for recent work)? etc.

Given the vast amount of possible string vacua, a promising strategy to address such questions would be to identify universal, model-independent constraints. These may help to rule out cosmological solutions in whole regions of the landscape and, even more importantly, point us to interesting regions where such solutions do exist.

A number of such no-go theorems is known in the literature, which constrain either the \emph{existence} or the \emph{stability} of dS solutions in certain corners of the landscape. Well-known examples of the first kind are the 11d/10d supergravity no-go theorems of \cite{Gibbons:1984kp, deWit:1986xg, Maldacena:2000mw} and the HKTT no-go theorem \cite{Hertzberg:2007wc}. In recent years, many more no-go theorems have been formulated for type II \cite{Silverstein:2007ac, Haque:2008jz, Steinhardt:2008nk, Caviezel:2008tf, Flauger:2008ad, Danielsson:2009ff, Caviezel:2009tu, Wrase:2010ew, VanRiet:2011yc, Andriot:2016xvq} and heterotic string theory \cite{Green:2011cn, Gautason:2012tb, Kutasov:2015eba, Quigley:2015jia}.

In the present paper, we are instead concerned with the (meta-)stability of given dS solutions. Due to the necessarily broken supersymmetry, one generically expects this to be an issue, and indeed tachyons are notoriously difficult to evade in many explicit string compactifications to lower-dimensional dS space-times. To some extent, this difficulty is explained by the fact that meta-stability is in general statistically unlikely \cite{Denef:2004cf, Marsh:2011aa, Chen:2011ac, Bachlechner:2012at}. In addition, however, there may also be structural reasons for the appearance of tachyons that are not captured by postulating a completely random supergravity potential.
For example, extensive scans of flux compactifications in the classical regime have discovered numerous dS critical points but not a single meta-stable one \cite{Caviezel:2008tf, Flauger:2008ad, Danielsson:2009ff, Caviezel:2009tu, Danielsson:2010bc, Danielsson:2011au}. On the other hand, constructions employing instanton effects or non-geometric fluxes are more successful in achieving meta-stability. It is of obvious importance to understand this, as unnecessary scans in inadequate corners of the landscape can then be avoided and one may focus on more promising ones instead.

Stability constraints have been addressed previously in several works, with a focus on those moduli that are universally present in all string compactifications. Constraints from the volume and dilaton moduli were worked out in \cite{Shiu:2011zt}, and the volume moduli associated to the cycles wrapped by O-planes were addressed in \cite{Danielsson:2012et}. Furthermore, it was argued in \cite{Covi:2008ea, Covi:2008cn} that the sgoldstino is the most dangerous modulus to become tachyonic near a supersymmetric/no-scale Minkowski point (see also \cite{Brustein:2004xn, GomezReino:2006dk, GomezReino:2006wv, GomezReino:2007qi, BenDayan:2008dv, Badziak:2008yg} for earlier work and \cite{Hetz:2016ics} for a recent extension to bending trajectories during inflation). In \cite{Junghans:2016uvg}, it was shown that the argument of \cite{Covi:2008ea} has to be generalized in the generic case where the mass matrix of the sgoldstino and the orthogonal moduli is non-diagonal. The tachyon is then not the sgoldstino but a different mode, which only aligns with the sgoldstino in the Minkowski limit. Based on explicit examples, it was furthermore conjectured in \cite{Junghans:2016uvg} that this mode is the reason why many compactifications do not admit meta-stable dS vacua even though they evade the known no-go theorems from the volume, the dilaton and the sgoldstino. It is the purpose of the present work to perform a general, model-independent analysis of this tachyon.

More precisely, we consider dS solutions which are obtained as a small deformation away from a no-scale Minkowski vacuum, e.g., by turning on additional fluxes or non-perturbative corrections in the superpotential. No-scale Minkowski vacua are abundant in the large-volume regime of string compactifications due to the form of the corresponding K{\"{a}}hler potential. Moreover, the vicinity of such points in moduli space is generally attractive since it may allow a controlled uplift of the vacuum energy. Indeed, moduli which are stabilized already at the Minkowski point are guaranteed to remain stabilized at the dS solution if the deformation is small enough. It is then sufficient to ensure that the originally massless moduli are stabilized as well. One therefore expects that meta-stable dS vacua are more likely to exist in the vicinity of such special points than at generic points in moduli space. Indeed, explicit scans suggest that the number of dS vacua rapidly drops to zero as one moves away from a Minkowski point (see, e.g., \cite{Danielsson:2012by}).

Specifically, we will consider string compactifications which admit a description in terms of 4d $\mathcal{N}=1$ supergravity with an F-term scalar potential and have a no-scale Minkowski vacuum somewhere in their parameter space. Our main result is that meta-stable dS vacua near such a no-scale Minkowski point are forbidden, unless perturbative string corrections to the K{\"{a}}hler potential are non-negligible, the superpotential contains a specific type of coupling between the no-scale and the perpendicular moduli, or at least one of the perpendicular moduli is unstabilized at the Minkowski point. We find that otherwise there is at least one tachyon in the spectrum, with second slow-roll parameter $\eta \le -\frac{4}{3}$. The tachyon is universal in the sense that its appearance can be proven model-independently. Its particular direction in moduli space is, however, model-dependent. The tachyon can be shown to align with the sgoldstino in the Minkowski limit but rotates away from it as one moves away from the Minkowski point. We also show that the sgoldstino itself can nevertheless be stable in the dS vacuum due to mixing effects in the mass matrix.

Apart from ruling out cosmological solutions in a large region of the landscape, our result may also help to explain why the program of finding classical dS vacua initiated in \cite{Hertzberg:2007wc, Silverstein:2007ac} remained unsuccessful \cite{Caviezel:2008tf, Flauger:2008ad, Danielsson:2009ff, Caviezel:2009tu, Danielsson:2010bc, Danielsson:2011au}. The idea there is to construct simple and explicit dS vacua from geometric flux compactifications including only minimal ingredients (i.e., NSNS, RR and metric fluxes as well as O-planes and possibly D-branes), but without employing perturbative or non-perturbative string corrections or ``exotic'' objects such as non-geometric fluxes. We argue that, in type IIA string theory, it is difficult---in some cases impossible---for such setups to evade our tachyon. Constructions involving instantons or non-geometric fluxes, on the other hand, can evade our no-go theorem, which is consistent with known examples using these ingredients.

This paper is organized as follows. In sections \ref{setup} and \ref{tachyon}, we spell out our assumptions and establish the no-go theorem. In section \ref{evade}, we discuss possibilities to evade the tachyon. In section \ref{class-dS}, we apply our results to classical dS solutions. We conclude in section \ref{concl} with a discussion of our results. Several longer computations are relegated to appendices \ref{app2}, \ref{app1} and \ref{stu-model}.

\section{Preliminaries}
\label{setup}

\subsection{Setup}

Our starting point is 4d $\mathcal{N}=1$ supergravity with an F-term scalar potential
\begin{equation}
V = \e^K \left( g^{I\bar J} D_I W D_{\bar J} \bar W - 3 |W|^2 \right) = \e^G \left( G_IG^I -3 \right) \label{fterm}
\end{equation}
in Planck units, where $D_I W = W_I + K_I W$ and $G=K+\ln |W|^2$. Indices are raised with the hermitian field space metric $g_{I \bar J}=K_{I \bar J}$. Here and in the following, we use subscripts to denote partial derivatives with respect to the $\Phi^I$ moduli (e.g., $K_I =\partial_I K$, $K_{I\bar J} = \partial_I\partial_{\bar J} K$, etc., and similar for $G$, $V$ and $W$).

For constant scalar fields, the equations of motion are
\begin{equation}
V_I = G_I V + \e^G G_I + \e^GG_{IJ}G^J - \e^GK_{IJ\bar K}G^JG^{\bar K} = 0 \label{eoms}
\end{equation}
and its complex conjugate. The components of the mass matrix are given by
\begin{align}
V_{I\bar J} &= K_{I \bar J} V + \e^G K_{I\bar J} + \e^G G_I G_{\bar J} + \e^G G_{IJ} G_{\bar J} G^J + \e^G G_{IJ} G_{\bar J}\vphantom{}^J - \e^G G_{I}\vphantom{}^{\bar K} K_{\bar J \bar K J} G^J \nl
 - \e^G K_{I J \bar K} G_{\bar J}\vphantom{}^{J}G^{\bar K} - \e^G K_{I J \bar K} G_{\bar J} G^J G^{\bar K} - \e^G K_{I \bar J J \bar K} G^J G^{\bar K} + \e^G K_{I J \bar K}K_{\bar J K}\vphantom{}^{J} G^K G^{\bar K} \nl + \e^G K_{I J \bar K}K_{\bar J \bar L}\vphantom{}^{\bar K} G^J G^{\bar L}, \label{massmatrix1} \\
V_{I J} &= G_{I J} V + 2 \e^G G_{I J} + \e^G G_I G_J + \e^G G_{IK} G_J G^K + \e^G G_{I J K} G^K - \e^G G_{I}\vphantom{}^{\bar K} K_{J \bar K K} G^K \nl
 - \e^G K_{I J \bar K} G^{\bar K} - \e^G K_{I K \bar K} G_J\vphantom{}^{\bar K}G^K - \e^G K_{I K \bar K} G_J G^K G^{\bar K} - \e^G K_{I J K \bar K} G^K G^{\bar K} \nl + \e^G K_{I L \bar K}K_{J K}\vphantom{}^L G^K G^{\bar K} + \e^G K_{I K \bar K}K_{J \bar L}\vphantom{}^{\bar K} G^K G^{\bar L} \label{massmatrix2}
\end{align}
and their complex conjugates, where we have used that the superpotential is holomorphic to simplify the expressions.

\subsection{No-scale Minkowski vacua}

Let us now consider Minkowski vacua of the potential \eqref{fterm} with a no-scale property \cite{Cremmer:1983bf, Ellis:1983sf}.
To this end, we split the moduli $\Phi^I$ into two sectors
\begin{equation}
\Phi^I = \left\{ \Phi^m, \Phi^a \right\},
\end{equation}
where we label the fields with the no-scale property \eqref{noscale-property} by $m,n,\ldots$. In the following, we will refer to them as no-scale moduli. The fields labelled by $a,b,\ldots$ denote possible orthogonal fields which are generically present in a given compactification and which we will assume to be stabilized at the Minkowski solution.

In order to admit a no-scale Minkowski vacuum, the K{\"{a}hler} potential and the superpotential need to satisfy the off-shell properties
\begin{equation}
K_mK^m = 3, \qquad W=W(\Phi^a). \label{noscale-property}
\end{equation}
Moreover, we assume that the K{\"{a}hler} potential separates and has an axionic shift symmetry as it is typical for bulk moduli in string compactifications,
\begin{equation}
K=K_1(\Phi^a+\bar \Phi^a)+K_2(\Phi^m+\bar\Phi^m).
\end{equation}
The scalar potential \eqref{fterm} then reduces to
\begin{equation}
V = \e^G G_aG^a.
\end{equation}
One can then check that, for $G_a=0$, all equations of motion are indeed satisfied at a Minkowski vacuum, $V=0$.

The components of the mass matrix at the Minkowski solution are
\begin{equation}
V_{\bar a b} = \e^G G_{\bar a \bar b}g^{\bar b c}G_{c b} + \e^G g_{\bar a b}, \qquad  V_{ab}=2\e^GG_{ab} \label{Vab}
\end{equation}
and $V_{mn}= V_{\bar m n} = V_{am}=V_{\bar a m} = 0$. Hence, the $\Phi^m$ moduli are massless, as expected from the no-scale property. The $\Phi^a$ moduli are stabilized if their mass matrix has no zero eigenvalues, which is equivalent to demanding
\begin{equation}
\det \begin{pmatrix}
   V_{\bar a}\vphantom{}^{\bar b} & V_{\bar a}\vphantom{}^{b} \\
   V_{a}\vphantom{}^{\bar b} & V_{a}\vphantom{}^{b}
\end{pmatrix} = \det \e^{G}\left({\begin{matrix}
   \delta_{\bar a}^{\bar b} & G_{\bar a}\vphantom{}^{b} \\
   G_{a}\vphantom{}^{\bar b} & \delta_a^b
   \end{matrix}}\right)^{\!2} \neq 0.
\end{equation}
Using that $\det \left(\begin{smallmatrix}
   A & B \\
   C & D
\end{smallmatrix}\right) = \det(AD-BC)$ for commuting matrices $C,D$, we arrive at the stability condition
\begin{equation}
\det\left(\delta_a^c - G_{ab}G^{bc}\right) \neq 0. \label{stab-cond}
\end{equation}

Let us finally mention that any non-trivial no-scale Minkowski vacuum in string theory must have a superpotential which is at least quadratic in some of the $\Phi^a$ moduli, i.e.,
\begin{equation}
W_{ab} \neq 0
\end{equation}
off-shell, where by non-trivial we mean that the superpotential is not identically zero. This property will be used in section \ref{class-dS} and is proven in appendix \ref{app2}.

\subsection{Assumptions}
\label{setup-assumptions}

In this paper, we will analyze the stability of dS solutions which are obtained as a deformation away from a no-scale Minkowski point in moduli space.
Specifically, we will make the following assumptions:
\begin{itemize}
\item $\e^K$ is a real homogeneous function of the moduli such that
\begin{equation}
K_m (\Phi^m + \bar \Phi^{\bar m}) = -3. \label{homog}
\end{equation}
This assumption is satisfied by string compactifications in the limit of large volume and small coupling, where $\alpha^\prime$ and $g_s$ corrections are negligible and the K{\"{a}}hler potential is of the usual log form. A simple example would be the potential $K = -3\ln(\Phi^1 +\bar \Phi^{\bar 1})-\ln(\Phi^2 +\bar \Phi^{\bar 2})-3\ln(\Phi^3 +\bar \Phi^{\bar 3})$, where either $\Phi^1$ oder $\Phi^3$ would play the role of the no-scale modulus $\Phi^m$, and the remaining two moduli would correspond to the $\Phi^a$. In the following, we will only assume the more general property \eqref{homog} to hold, which captures models with an arbitrary number of moduli. Also note that the K{\"{a}}hler potential need in general not be a sum of terms involving only one modulus each. Instead, the above assumption is also satisfied, e.g., for Calabi-Yau compactifications with general intersection numbers. In section \ref{evade}, we will consider more general K{\"{a}}hler potentials violating \eqref{homog} and see that our no-go theorem can be evaded if perturbative string corrections are large enough.

Eq. \eqref{homog} implies the useful identities
\begin{equation}
K_mK^m = 3 \label{contractions1}
\end{equation}
and
\begin{align}
& K^m K_{mn} = K_{n}, && K^m K_{mnl} = 2 K_{nl}, && K^m K_{mnlp}  = 3 K_{nlp}, && K^{ml}K_{ln}=\delta_n^m. \label{contractions2}
\end{align}
Analogous identities hold for contractions involving the barred indices. Note that mixed components such as $K_{am}$ vanish. Also note that all of the above identities for $K$ and its derivatives hold off-shell.
\item The model has a no-scale Minkowski point somewhere in parameter space,
\begin{equation}
K_mK^m = 3, \qquad \lim_{\lambda \to 0} W(\Phi^a,\Phi^m;\lambda) = W(\Phi^a), \label{assumption}
\end{equation}
at which all $\Phi^a$ moduli are stabilized, cf. \eqref{stab-cond}. Here, $\lambda$ is a deformation parameter which we will explain momentarily.

Assumption \eqref{assumption} is typically true for flux compactifications in the large-volume regime or at large complex structure. For particular choices of the flux numbers, they are known to admit Minkowski vacua, which indeed have a no-scale structure (see, e.g., \cite{Giddings:2001yu, Camara:2007cz, Palti:2008mg, Blaback:2010sj, Andriot:2016ufg} for examples). As one turns on additional fluxes and/or non-perturbative corrections in the superpotential, the no-scale structure can be broken such that the resulting scalar potential has AdS or dS extrema. We parametrize such a general deformation by a parameter $\lambda$, which we define such that the Minkowski point is located at $\lambda=0$.

As stated above, also corrections to the K{\"{a}}hler potential can break the no-scale structure, but for the moment we will assume that this is not the case and defer a discussion of more general K{\"{a}}hler potentials to section \ref{evade}.
In other words, at finite $\lambda$, the K{\"{a}}hler potential is still of the above form and satisfies the off-shell identities \eqref{homog}--\eqref{contractions2}, while the superpotential can acquire a $\Phi^m$ dependence.
\item The dS solution is close to the no-scale Minkowski vacuum in the sense that
\begin{equation}
\lambda \ll 1. \label{small-lambda}
\end{equation}

We can then perform a systematic expansion of the on-shell values of the superpotential and the K{\"{a}}hler potential and their derivatives around the Minkowski point. As stated above, the off-shell superpotential receives a $\lambda$-dependence by turning on combinations of fluxes and/or non-perturbative corrections. Through the equations of motion, the $\lambda$-deformation then back-reacts on the vevs of the scalar fields and, hence, on the on-shell values of the superpotential and the K{\"{a}}hler potential and their derivatives. Hence, we can write
\begin{equation}
W =  W^{(0)} + \lambda W^{(1)} + \mathcal{O}(\lambda^2)
\end{equation}
for the on-shell superpotential. Using that, at the Minkowski solution for $\lambda=0$, $W_m=W_{am}=W_{mn}=0$ and $G_a=0$, the on-shell values of the derivatives of the superpotential can furthermore be expanded as
\begin{align}
& W_a =-K^{(0)}_a W^{(0)} + \mathcal{O}(\lambda), && W_{ab} = W_{ab}^{(0)} + \mathcal{O}(\lambda), && W_{am} = \lambda W_{am}^{(1)} + \mathcal{O}(\lambda^2), \notag \\ & W_m = \lambda W_m^{(1)} + \mathcal{O}(\lambda^2), && W_{mn} = \lambda W_{mn}^{(1)} + \mathcal{O}(\lambda^2), && \ldots
\end{align}
and analogously for higher derivatives.\footnote{In order to avoid confusion, note that $W_m^{(1)}$ in our notation is the first order coefficient in the expansion of the on-shell value of $W_m$, which is not the same as $\partial_m (W^{(1)})$ (and analogous for the other coefficients).}
Similarly, the on-shell K{\"{a}}hler potential and its derivatives can be expanded as
\begin{equation}
K = K^{(0)} + \mathcal{O}(\lambda), \quad K_a = K^{(0)}_a + \mathcal{O}(\lambda), \quad K_m = K^{(0)}_m + \mathcal{O}(\lambda), \quad \ldots
\end{equation}
etc. Note that, by assumption, the contraction identities \eqref{contractions1} and \eqref{contractions2} hold to all orders in the $\lambda$-expansion (on-shell as well as off-shell).

Finally, let us make a remark about assumption \eqref{small-lambda}. In general, the $\lambda$-deformation can be generated by an arbitrary combination of ingredients to $W$ breaking the no-scale structure, such as extra fluxes and non-perturbative effects. In the special case where the $\lambda$-deformation is purely generated by turning on additional fluxes, one might wonder whether flux quantization might be an obstruction to making $\lambda$ small. Two comments are in order here. First, we regard the issue of flux quantization as an \emph{additional} obstacle that a consistent dS vacuum needs to surpass, on top of solving the equations of motion and having a positive definite mass matrix. What we show in this paper is that, even ignoring this additional difficulty, there is already a no-go theorem for a large class of string compactifications. 
Second, even if $\lambda \gtrsim \mathcal{O}(1)$ due to quantized fluxes in a given model, the corresponding dS vacuum may still be close to a no-scale Minkowski point if $W^{(0)} \gg \lambda W^{(1)}$. This is the case, for example, when the flux numbers generating $W^{(0)}$ are large. Rescaling the full superpotential by an overall factor (which leaves the equations of motion invariant), one can map such a setup to one with $\lambda \ll 1$. Hence, our arguments then still apply.
\item The superpotential satisfies
\begin{equation}
W_{amn}^{(1)} = 0 \label{wamn}
\end{equation}
on-shell. This is in particular true if mixed couplings between the no-scale moduli and the $\Phi^a$ moduli generating the above term are absent in the superpotential. As we will discuss in section \ref{class-dS}, this is indeed the case in type IIA models where the superpotential is purely generated by geometric fluxes. Also note that, in models where the potential for the no-scale directions is generated by non-perturbative terms, it is often assumed that the $\Phi^a$-dependence of the one-loop determinant is negligible. In such models, the above term then vanishes as well. In a general situation including all types of fluxes and non-perturbative terms, however, $W_{amn}$ can be non-zero and relevant. In section \ref{evade}, we will relax the assumption \eqref{wamn} and see that this can help to evade our no-go theorem if $W_{amn}$ is sufficiently large.
\end{itemize}

\section{A universal tachyon}
\label{tachyon}

In order to establish our no-go theorem, we will now compute the mass of a field, $\Psi$, which we define as the combination
\begin{equation}
\Psi := T_I\Phi^I, \qquad \Phi^I = \frac{T^I \Psi}{T_JT^J}. \label{ti1}
\end{equation}
Here, $T_I$ is a vector in field space given by
\begin{equation}
T_I = \delta^m_I K_m
+ \delta^a_I Y_a \label{ti3}
\end{equation}
with a priori undetermined
$Y_a$, which we define such that it goes to zero in the limit $\lambda\to 0$, i.e.,
\begin{equation}
Y_a = \lambda Y_a^{(1)} + \mathcal{O}(\lambda^2).
\end{equation}
Substituting the $\lambda$-expansion into \eqref{ti3}, we thus have
\begin{equation}
T_I = \delta^m_I K_m^{(0)} + \lambda \delta^m_I K_m^{(1)} +
\lambda \delta^a_I Y_a^{(1)} + \mathcal{O}(\lambda^2). \label{ti2}
\end{equation}

The masses of the real and imaginary parts of $\Psi$ after canonical normalization are
\begin{align}
& m_{\text{Re}\Psi}^2 = \frac{1}{2T_IT^I}\left( V_{\bar IJ}T^{\bar I} T^J + V_{I\bar J}T^I T^{\bar J}+ V_{IJ}T^I T^J + V_{\bar I\bar J}T^{\bar I} T^{\bar J} \right), \label{psimass1} \\
& m_{\text{Im}\Psi}^2 = \frac{1}{2T_IT^I}\left( V_{\bar IJ}T^{\bar I} T^J + V_{I\bar J}T^I T^{\bar J} - V_{IJ}T^I T^J - V_{\bar I\bar J}T^{\bar I} T^{\bar J} \right). \label{xy}
\end{align}
As we will see momentarily, $\text{Re}\Psi$ and $\text{Im}\Psi$ imply the existence of a tachyon if $T_I$ (i.e., $Y_a$) is chosen appropriately.
\\

\subsection{ Case 1: $W_{mn}=W_{mnr}=0$}

Let us first consider the case where the superpotential is linear in the $\Phi^m$ moduli or, more generally, satisfies
\begin{equation}
W_{mn}=W_{mnr}=0
\end{equation}
at the dS solution. As we will discuss in more detail in section \ref{class-dS}, this is the case relevant, e.g., for many geometric flux compactifications. Note that we do not make any assumption about the $\Phi^a$ dependence of the superpotential, except that the $\Phi^a$ moduli are all stabilized at leading order $\mathcal{O}(\lambda^0)$.

Substituting \eqref{massmatrix1}, \eqref{massmatrix2} and \eqref{ti3} into \eqref{psimass1}, we find that, up to quadratic order in $\lambda$,
\begin{align}
m_{\text{Re}\Psi}^2 &= -\frac{4}{3}V + \frac{1}{3} \e^{G} \left| G_a + Y_a - G_{ab} \left( G^{b} - \bar Y^{b} \right) \right|^2 + \mathcal{O}(\lambda^3), \label{tachyon-mass}
\end{align}
where we repeatedly used the equations of motion and the identities \eqref{contractions1} and \eqref{contractions2} to simplify the expression. Since the computation is rather tedious, we have relegated the details to appendix \ref{app1} for the interested reader.\footnote{The computation can conveniently be done using the Mathematica xTensor package \url{http://www.xact.es/xTensor/}.} Analogous manipulations in \eqref{xy} furthermore show that the mass of the imaginary part of $\Psi$ is manifestly non-negative at the quadratic order,
\begin{align}
m_{\text{Im}\Psi}^2 &= \frac{1}{3}  \e^G \left| G_a + Y_a + G_{ab} \left( G^b - \bar Y^b \right) \right|^2 + \mathcal{O}(\lambda^3). \label{impsi-mass}
\end{align}
Note that the squared terms in \eqref{tachyon-mass} and \eqref{impsi-mass} differ by a relative sign. In order not to clutter the equations with indices, we have refrained from explicitly substituting the $\lambda$-expansion in \eqref{tachyon-mass} and \eqref{impsi-mass}. However, one can check that each term on the right-hand sides of both equations is of the order $\lambda^2$. For the squares $\left| \ldots \right|^2$, this follows from $G_a \sim Y_a \sim \lambda$. The on-shell scalar potential, on the other hand, is naively of the order $\lambda$ since
\begin{equation}
V = \e^G \left( K_m \frac{\bar W^{m}}{\bar W} + K^m \frac{W_m}{W} \right) + \mathcal{O}(\lambda^2).
\end{equation}
However, the equations of motion \eqref{eoms} imply that this vanishes at linear order:
\begin{equation}
0 = K^m V_m = 2\e^G \left( K_m \frac{\bar W^{m}}{\bar W} + K^m \frac{W_m}{W} \right) + \mathcal{O}(\lambda^2). \label{kmvm}
\end{equation}
Note that this does in general not mean that $W_m^{(1)}=0$ but only that the particular combination in \eqref{kmvm} is of order $\lambda^2$.

Hence, if we can find a $Y_a$ minimizing the square on the right-hand side of \eqref{tachyon-mass},
\begin{equation}
\exists Y_a: \quad G_a + Y_a - G_{ab} \left( G^b - \bar Y^b \right)=0, \label{yeq1}
\end{equation}
then $m_{\text{Re}\Psi}^2$ is proportional to the on-shell scalar potential. By substituting \eqref{yeq1} back into itself, we can decouple $Y_a$ from its complex conjugate and arrive at
\begin{equation}
Y_a - G_{ab}G^{bc}Y_c + G_a-2G_{ab}G^b+G_{ab}G^{bc}G_c = 0. \label{yeq2}
\end{equation}
A system of linear equations such as \eqref{yeq2} always has a solution provided that the determinant of the coefficient matrix is non-zero, $\det \left( \delta_a^c - G_{ab}G^{bc} \right) \neq 0$. Recalling \eqref{stab-cond}, this is indeed true whenever all $\Phi^a$ moduli are stabilized at leading order.

Hence, one can always choose a $Y_a$ such that the square on the right-hand side of \eqref{tachyon-mass} vanishes. This implies that, at any dS critical point, $V>0$, the mass matrix contains a tachyon with
\begin{equation}
\eta \le -\frac{4}{3}, \label{eta}
\end{equation}
where $\eta$ is the second slow-roll parameter. The direction of the tachyon in field space depends on the solution $Y_a$ in \eqref{yeq1} and, hence, on how the $\Phi^a$ moduli are stabilized. It is therefore model-dependent. The appearance of the tachyon is, however, universal. It is important to note that it does not matter whether $\text{Re}\Psi$ is an eigenmode of the mass matrix. Any direction with a negative $m^2$ implies the existence of at least one negative eigenvalue of the full mass matrix. This is a consequence of Sylvester's criterion.

Note that \eqref{stab-cond} implies that also the square in \eqref{impsi-mass} can be set to zero for a certain choice of $Y_a$ (which is in general different from the $Y_a$ that sets the square in \eqref{tachyon-mass} to zero). This again follows from solving a system of linear equations for $Y_a$ and demanding that the determinant of the coefficient matrix is non-zero. At least one combination of the moduli therefore remains massless at $\mathcal{O}(\lambda^2)$ if all $\Phi^a$ moduli are stabilized at leading order.

We should furthermore stress that, under our assumptions, the above result also excludes slow-roll inflation. Away from a critical point $V_I=0$, the on-shell identity \eqref{tachyon-mass} receives corrections proportional to $V_I$ such that \eqref{eta} becomes $\eta \le -\frac{4}{3} + \mathcal{O}(\sqrt{\epsilon})$, with $\epsilon$ the first slow-roll parameter. We thus cannot have both $\epsilon$ and $|\eta|$ small and sustained slow-roll is not possible.

Let us now compare this to the sgoldstino mass. The sgoldstino $S$ is defined as the direction in field space along which supersymmetry is broken,
\begin{equation}
S = G_I\Phi^I, \qquad \Phi^I = \frac{G^I S}{G_JG^J}.
\end{equation}
The vector $G_I$ can be written as
\begin{equation}
G_I = \delta^m_I K_m + \delta^m_I \frac{W_m}{W} + \delta^a_I G_a,
\end{equation}
where we remind the reader that $\frac{W_m}{W} \sim \lambda$, $G_a \sim \lambda$. Comparing this to \eqref{ti1} and \eqref{ti2}, we thus conclude that the tachyon aligns with the sgoldstino in the limit $\lambda \to 0$. One furthermore checks that the sgoldstino mass is (cf. appendix \ref{app1})
\begin{align}
m_{\text{Re}S}^2 = -\frac{4}{3}V + \frac{4}{3} \e^G G_aG^a + \mathcal{O}(\lambda^3), \qquad m_{\text{Im}S}^2 &= \frac{4}{3}  \e^G G_aG^a + \mathcal{O}(\lambda^3).
\end{align}
The sgoldstino itself is thus stable except in the special case where SUSY breaking along the $\Phi^a$ directions is zero or sufficiently small. This phenomenon is indeed realized in explicit models \cite{Junghans:2016uvg} and explained by mass mixing effects as follows. At the Minkowski point, the no-scale moduli are massless, while the $\Phi^a$ moduli are stabilized.
One may then suspect that only the no-scale moduli are in danger of being destabilized by a small $\lambda$-deformation. In fact, the only vector one can construct model-independently at $\mathcal{O}(\lambda^0)$ is $K_m$ such that the only model-independent combination of the no-scale moduli is the sgoldstino. According to this logic, a universal tachyon, if existent, could only be the sgoldstino. However, in a general model, the $\lambda$-deformation does not only enter in the mass terms of the $\Phi^m$ and $\Phi^a$ moduli but also in off-diagonal terms in the mass matrix. It can then happen that the sgoldstino itself is stabilized by the deformation, while a particular combination of the $\Phi^m$ and $\Phi^a$ moduli is destabilized.
As we have just shown, this is precisely what happens in any string compactification with $\e^G G_aG^a > V$ (as, e.g., in \cite{Junghans:2016uvg}).

Finally, note that, as $\lambda$ is increased beyond the regime $\lambda \ll 1$, higher order corrections to the leading order tachyon mass become relevant. In order to still infer $\eta \le -\frac{4}{3}$, one then needs to take into account higher order corrections to the field direction defined by $T_I$ such that the property $m_{\text{Re}\Psi}^2 = -\frac{4}{3} V$ remains intact. Such corrections were successfully computed in \cite{Junghans:2016uvg} for a particular flux compactification of type IIA string theory. We leave a general derivation of such corrections for future work. Furthermore, for large enough $\lambda$, the $\lambda$-expansion may break down altogether. It would be very interesting if one could find an all-order resummation of our result that is also valid outside the convergence radius of the $\lambda$-expansion.

\subsection{Case 2: $W_{mn}\neq 0, W_{mnr}=0$}

We now discuss what happens when we allow the superpotential to be quadratic in the no-scale moduli, i.e.,
\begin{equation}
W_{mn} \neq 0, \qquad W_{mnr}=0
\end{equation}
at the dS solution. In that case, we find the leading order result
\begin{equation}
m_{\text{Re}\Psi}^2 = -\frac{4}{3}\e^G \left( K_m \frac{\bar W^{m}}{\bar W} + K^m \frac{W_m}{W} \right) + \mathcal{O}(\lambda^2). \label{x1}
\end{equation}
Unlike above, the right-hand side does not vanish at linear order by the equations of motion anymore since now
\begin{equation}
0 = K^m V_m = \e^G \left( 2K_m \frac{\bar W^{m}}{\bar W} + 2 K^m \frac{W_m}{W} + K^m K^n \frac{W_{mn}}{W} \right) + \mathcal{O}(\lambda^2). \label{x2}
\end{equation}
Rewriting $m_{\text{Re}\Psi}^2$ in terms of the on-shell scalar potential, we find
\begin{equation}
m_{\text{Re}\Psi}^2 = -\frac{4}{3} V + \mathcal{O}(\lambda^2)
\end{equation}
such that we again have a tachyon with $\eta \le -\frac{4}{3}$. Note that, unlike for the case $W_{mn}=0$, the leading order tachyon mass is not sensitive to the subleading correction $Y_a$ to the tachyon direction. This implies that the tachyon is present independently of whether the $\Phi^a$ moduli are stabilized at leading order or not.

A special case occurs when $K^m K^n \frac{W_{mn}}{W}$ vanishes on-shell at linear order in $\lambda$, which happens if $W_{mn}^{(1)}=0$ or, more generally, $K^{m(0)}K^{n(0)}W_{mn}^{(1)}=0$. By \eqref{x1} and \eqref{x2}, one then has $m_{\text{Re}\Psi}^2 \sim \mathcal{O}(\lambda^2)$. This case is a bit more involved and discussed in full generality in the next subsection.

\subsection{Case 3: $W_{mnr}\neq 0$}
\label{case3}

Let us now consider a completely general $\Phi^m$-dependence in $W$ in the sense that we also allow
\begin{equation}
W_{mnr} \neq 0
\end{equation}
along with $W_{mn}\neq 0$ at the dS solution. As the only remaining restriction, we do not yet allow $W_{amn}^{(1)}\neq 0$ (as explained in section \ref{setup-assumptions}).
Note that at most third derivatives of $W$ appear in the mass matrix.
In order for the dependence of the superpotential on the no-scale moduli to vanish at the Minkowski point, 
we require that $W_{mnr}$ is non-zero earliest at linear order in the $\lambda$-expansion, $W_{mnr} = \lambda W_{mnr}^{(1)} + \mathcal{O}(\lambda^2)$.

We now consider the mass of a specific combination of $\text{Re}\Psi$ and $\text{Im}\Psi$, i.e., the field $\text{Re}(\e^{-i\varphi/2}\Psi) = \cos(\varphi/2) \text{Re}\Psi + \sin(\varphi/2) \text{Im}\Psi$. The mass, $m^2$, of this field can be computed 
analogously to $m_{\text{Re}\Psi}^2$ and $m_{\text{Im}\Psi}^2$ and reduces to these masses for, respectively, $\varphi=0$ and $\varphi=\pi$. For the details of the mass computation, we again refer to appendix
\ref{app1}.

We distinguish four cases:

\subsubsection{Case 3a: $K^m K^n\frac{W_{mn}}{W}=\mathcal{O}(\lambda)$, $K^mK^nK^p\frac{W_{mnp}}{W}=\mathcal{O}(\lambda)$}

Assuming $K^m K^n\frac{W_{mn}}{W}=\mathcal{O}(\lambda)$, $V$ is non-vanishing already at linear order in $\lambda$, and the mass at this order is
\begin{align}
m^2 &= -\frac{2}{3}\left(1+\cos\varphi\right)V + \frac{1}{6} \e^G \left( \e^{-i\varphi} K_mK_nK_r \frac{\bar W^{mnr}}{\bar W} + \e^{i\varphi} K^mK^nK^r \frac{W_{mnr}}{W} \right) \nl + \mathcal{O}(\lambda^2). \label{m2-case3l}
\end{align}
Defining
\begin{equation}
K^mK^nK^r\frac{W_{mnr}}{W} = |w| \e^{i\delta},
\end{equation}
this becomes
\begin{align}
m^2 &= -\frac{2}{3}\left(1+\cos\varphi\right)V + \frac{1}{3} \e^G |w| \cos (\varphi+\delta) + \mathcal{O}(\lambda^2).
\end{align}
The first term on the right-hand side is proportional to $V$ and thus negative for all $ \varphi\neq \pi$ at a dS solution at linear order in $\lambda$. 
The second term, on the other hand, is
periodic and necessarily negative for a finite range of $\varphi$-values. Hence, there is always a linear combination of $\text{Re}\Psi$ and $\text{Im}\Psi$ which is 
tachyonic at linear order in $\lambda$.

\subsubsection{Case 3b: $K^m K^n\frac{W_{mn}}{W}=\mathcal{O}(\lambda)$, $K^mK^nK^p\frac{W_{mnp}}{W}=\mathcal{O}(\lambda^2)$}

In this case, the potential $V$ at the de Sitter solution is still of linear order in $\lambda$, but the second term in (\ref{m2-case3l}) is subleading, so that we always
have a tachyon
at linear order in $\lambda$ for for any choice $\varphi\neq \pi$, just as in case 2.

\subsubsection{Case 3c: $K^m K^n\frac{W_{mn}}{W}=\mathcal{O}(\lambda^2)$, $K^mK^nK^p\frac{W_{mnp}}{W}=\mathcal{O}(\lambda)$}

In this case, $V$ is at least of quadratic order in $\lambda$ so that the second term in (\ref{m2-case3l}) dominates and always 
gives rise to a tachyon at linear order in $\lambda$, as one can always choose $\varphi$ such that $\cos(\varphi+\delta)$ is negative. 

\subsubsection{Case 3d: $K^m K^n\frac{W_{mn}}{W}=\mathcal{O}(\lambda^2)$, $K^mK^nK^p\frac{W_{mnp}}{W}=\mathcal{O}(\lambda^2)$}

A special case occurs when both $K^mK^nK^r \frac{W_{mnr}}{W}$ and $K^m K^n \frac{W_{mn}}{W}$ vanish at linear order in the $\lambda$-expansion, 
i.e., $K^mK^nK^r \frac{W_{mnr}}{W}=\mathcal{O}(\lambda^2)$ and $K^m K^n \frac{W_{mn}}{W}=\mathcal{O}(\lambda^2)$. 
As discussed previously, the latter implies $V =\mathcal{O}(\lambda^2)$ by the equations of motion. The leading terms in $m^2$ are therefore now quadratic in $\lambda$,
\begin{align}
m^2 &= -\frac{2}{3}\left(1+\cos\varphi\right)V + \frac{1}{6} \e^G \left( \e^{-i\varphi} K_mK_nK_r \frac{\bar W^{mnr}}{\bar W} + \e^{i\varphi} K^mK^nK^r \frac{W_{mnr}}{W} \right) \nl +  \frac{1}{6}\e^{G}\left( \e^{-i\varphi} \frac{\bar W^{mnr}K_m K_nW_r}{W\bar W} + \e^{i\varphi} \frac{W_{mnr}K^mK^n\bar W^r}{W\bar W}\right) \nl
- \frac{1}{3}\e^{G}\left( \e^{-i\varphi} \frac{\bar W^{mn}K_m W_n}{W\bar W} + \e^{i\varphi} \frac{W_{mn}K^m\bar W^n}{W\bar W} \right) \nl 
+ \frac{1}{3} \e^{G} \left| G_a + Y_a - \e^{i\varphi} G_{ab} \left(G^b - \bar Y^b\right) \right|^2 + \mathcal{O}(\lambda^3). \label{m2-case3}
\end{align}
The last term on the right-hand side vanishes if we choose $Y_a$ appropriately. A tachyon then follows again from the same argument that we made above. We first write
\begin{equation}
K^mK^n\left(K^r+ \frac{\bar W^r}{\bar W}\right)\frac{W_{mnr}}{W} - 2 \frac{W_{mn}K^m\bar W^n}{W\bar W} = |w| \e^{i\delta}.
\end{equation}
We thus find
\begin{align}
m^2 &= -\frac{2}{3}\left(1+\cos\varphi\right)V + \frac{1}{3} \e^G |w| \cos (\varphi+\delta) + \mathcal{O}(\lambda^3).
\end{align}
The first term on the right-hand side is proportional to $V$ and thus negative for all $\varphi\neq\pi$ at a dS solution. The second term, on the other hand, is either identically zero (for $|w|=0$) or periodic and necessarily negative for a certain $\varphi$. Hence, there is always a linear combination of $\text{Re}\Psi$ and $\text{Im}\Psi$ which is tachyonic at quadratic order in $\lambda$.
\\

To conclude this section, we discuss a few special cases of our no-go theorem that were addressed in the literature before. We first consider a model where supersymmetry breaking only happens along the no-scale direction, i.e., $G_a=0$. The square in \eqref{tachyon-mass} can then be set to zero by choosing $Y_a=0$, independent of whether the $\Phi^a$ moduli are stabilized at leading order. One thus has a tachyon along the sgoldstino direction such that the argument of \cite{Covi:2008ea} applies. As a second example, consider a model with a single no-scale modulus and a quadratic superpotential. Our discussion above then again implies a tachyon, which reproduces a no-go theorem in \cite{Brustein:2004xn}. Finally, consider a model with one no-scale modulus and a superpotential such that $W_m^{(1)}=W_{mn}^{(1)}=W_{am}^{(1)}=\ldots=0$. We then recover the scenario studied in \cite{Marsh:2014nla}, which was argued to admit meta-stable dS vacua upon a small tuning of parameters. The equations of motion $V_a=0$ then imply $G_a+G_{ab}G^b=\mathcal{O}(\lambda^2)$ and, hence, $G_a -G_{ab}G^{bc}G_c = \mathcal{O}(\lambda^2)$. Thus, the matrix $\delta_a^c - G_{ab}G^{bc}$ has at least one zero eigenvalue at leading order, i.e., at least one of the $\Phi^a$ is not stabilized at leading order. The scenario of \cite{Marsh:2014nla} thus consistently evades our no-go theorem.
\\

Let us summarize this section. We have shown that, under the assumptions spelled out above, it is not possible to obtain meta-stable dS vacua or solutions suitable for inflation close to a no-scale Minkowski solution. If the superpotential is linear in the no-scale moduli or, more generally, satisfies $K^mK^nK^r \frac{W_{mnr}}{W}=\mathcal{O}(\lambda^2)$ and $K^m K^n \frac{W_{mn}}{W}=\mathcal{O}(\lambda^2)$, the tachyon is present whenever the $\Phi^a$ moduli are stabilized at leading order, where its specific direction in field space is model-dependent. The tachyon aligns with the sgoldstino in the Minkowski limit, while the sgoldstino itself can be stable in the dS vacuum. If, on the other hand, the superpotential has a general dependence on the no-scale moduli with $K^mK^nK^r \frac{W_{mnr}}{W}=\mathcal{O}(\lambda)$ and/or $K^m K^n \frac{W_{mn}}{W}=\mathcal{O}(\lambda)$, the tachyon is present independently of whether the $\Phi^a$ moduli are stabilized or not at leading order.

\section{Evading the no-go theorem}
\label{evade}

In this section, we will discuss the possibilities to evade our no-go theorem under the assumption that the scalar potential is still a pure F-term scalar potential and that the dS vacuum is close to a no-scale Minkowski vacuum. Specifically, we will show that at least one of the following three necessary conditions then needs to be satisfied:
\begin{itemize}
 \item The K{\"{a}}hler potential receives sizable perturbative corrections breaking the no-scale structure.
 \item The superpotential has couplings such that $W_{amn}$ is non-zero at linear order in the $\lambda$-expansion.
 \item At least one (linear combination) of the $\Phi^a$ moduli is not stabilized at leading order.
 Here and in the following, by ``unstabilized $\Phi^a$'' we mean that either $\text{Re}\Phi^a$ or $\text{Im}\Phi^a$ or a linear combination of them is not stabilized at leading order, while the orthogonal combination is stabilized. One can check using \eqref{Vab} that, for non-zero $\e^G$, it is not possible to have both $\text{Re}\Phi^a$ and $\text{Im}\Phi^a$ unstabilized.
\end{itemize}
For simplicity, we will only consider each of these possibilities individually as this is sufficient to evade our no-go theorem.

\subsection{Corrections to the K{\"{a}}hler potential}
\label{evade2}

We now study how the tachyon mass is affected by $\alpha^\prime$ or $g_s$ corrections to the leading K{\"{a}}hler potential. We consider a small correction $k$ of the form
\begin{align}
\hat K &= K_1(\Phi^a+\bar\Phi^a) + K_2(\Phi^m+\bar\Phi^m) + k(\Phi^a,\bar\Phi^a,\Phi^m,\bar\Phi^m),
\end{align}
where we denote the corrected K{\"{a}}hler potential by $\hat K$ and the leading potential is given by $K=K_1+K_2$ satisfying \eqref{homog}. Note that $k$ can in general depend on both the $\Phi^a$ and the $\Phi^m$ moduli such that it generates non-zero mixed components $g_{\bar a m}$ of the field space metric.

Repeating our computation of the $\Psi$ mass, we find that it receives a correction
\begin{align}
\Delta m_{\text{Re}\Psi}^2 &= \frac{1}{6} \e^G \left( k_{mnr}K^mK^nK^r + 7 k_{\bar m nr}K^{\bar m}K^nK^r + 7 k_{\bar m \bar nr}K^{\bar m}K^{\bar n}K^r + k_{\bar m \bar n\bar r}K^{\bar m}K^{\bar n}K^{\bar r}\right) \nl - \frac{1}{6}\e^G \left( k_{\bar m nr s}K^{\bar m}K^nK^rK^s + 2 k_{\bar m \bar nrs}K^{\bar m}K^{\bar n}K^rK^s + k_{\bar m \bar n\bar r s}K^{\bar m}K^{\bar n}K^{\bar r}K^s \right) \nl + \mathcal{O}(k^2,k\lambda), \\
\Delta m_{\text{Im}\Psi}^2 &= - \frac{1}{6} \e^G \left( k_{mnr}K^mK^nK^r - k_{\bar m nr}K^{\bar m}K^nK^r - k_{\bar m \bar nr}K^{\bar m}K^{\bar n}K^r + k_{\bar m \bar n\bar r}K^{\bar m}K^{\bar n}K^{\bar r}\right) \nl + \frac{1}{6}\e^G \left( k_{\bar m nr s}K^{\bar m}K^nK^rK^s - 2 k_{\bar m \bar nrs}K^{\bar m}K^{\bar n}K^rK^s + k_{\bar m \bar n\bar r s}K^{\bar m}K^{\bar n}K^{\bar r}K^s \right) \nl + \mathcal{O}(k^2,k\lambda),
\end{align}
where for simplicity we have only written down corrections linear in $k$ and leading in $\lambda$. Note that only derivatives with respect to the $\Phi^m$ moduli appear at the above leading order.

For a correction which respects the shift symmetry of the axionic parts of the moduli, we can write
\begin{equation}
k=k(\Phi^a+\bar\Phi^a,\Phi^m+\bar\Phi^m). \label{shift-symmetric}
\end{equation}
This should be true for any perturbative $\alpha^\prime$ or $g_s$ correction to $K$.
The corrections to $m_{\text{Re}\Psi}^2$ and $m_{\text{Im}\Psi}^2$ then simplify,
\begin{align}
\Delta m_{\text{Re}\Psi}^2 &= \frac{8}{3} \e^G k_{mnr}K^mK^nK^r - \frac{2}{3}\e^G k_{m nr s}K^{m}K^nK^rK^s + \mathcal{O}(k^2,k\lambda), \label{psi-mass-k} \\
\Delta m_{\text{Im}\Psi}^2 &= \mathcal{O}(k^2,k\lambda).
\end{align}
Hence, $k$ only helps to stabilize $\text{Re}\Psi$ but not $\text{Im}\Psi$, as expected from \eqref{shift-symmetric}.

So far, we did not make any assumption about the particular form of the correction $k$.
For the sake of an explicit example, we will now assume that the correction takes the form
\begin{equation}
k(\Phi^a,\bar\Phi^a,\Phi^m,\bar\Phi^m) = -\xi(\Phi^a,\bar\Phi^a) \e^{K_2(\Phi^m+\bar\Phi^m)/2}. \label{def-k}
\end{equation}
A well-known example for a correction of the above type is the $\mathcal{O}(\alpha^{\prime 3})$ BBHL correction to Calabi-Yau orientifold compactifications of type IIB \cite{Becker:2002nn} (see also \cite{Bonetti:2016dqh}). The no-scale moduli $\Phi^m$ are then the K{\"{a}}hler moduli of the Calabi-Yau, and
\begin{equation}
\xi = -\frac{\chi(\Sigma)\zeta(3)}{2} \e^{-3\phi_0/2}, \qquad K_2 = -2 \ln \mathcal{V}.
\end{equation}
Here, $\phi_0$ is the dilaton, $\chi(\Sigma)$ is the Euler characteristic of the Calabi-Yau, and $\mathcal{V}$ is its classical volume, which implicitly depends on the K{\"{a}}hler moduli.

From \eqref{def-k}, \eqref{contractions1} and \eqref{contractions2}, we can derive the identities
\begin{equation}
k_{mnr}K^mK^nK^r = \frac{105}{8}k, \quad k_{mnrs}K^mK^nK^rK^s = \frac{945}{16}k
\end{equation}
and analogous for contractions involving barred indices. The correction to the mass term of $\text{Re}\Psi$ thus becomes
\begin{align}
\Delta m_{\text{Re}\Psi}^2 &= - \frac{35}{8}\e^G k + \mathcal{O}(k^2,k\lambda).
\end{align}
Depending on the how large the correction $k$ is and which sign it has, it may therefore stabilize the tachyon.
The correction has a stabilizing effect whenever $\xi > 0$. For the case of a IIB Calabi-Yau orientifold, this corresponds to a negative Euler characteristic. This is indeed assumed in the familiar large-volume scenario \cite{Balasubramanian:2005zx} (which in addition utilizes the effect of non-perturbative corrections to the superpotential and an uplifting term).

As a second example, let us consider a correction of the form
\begin{equation}
k(\Phi^a,\bar\Phi^a,\Phi^m,\bar\Phi^m) = -\xi(\Phi^a,\bar\Phi^a) \e^{K_2(\Phi^m+\bar\Phi^m)/3}.
\end{equation}
Note the different factor 3 in the exponential compared to \eqref{def-k}.
We then find
\begin{equation}
k_{mnr}K^mK^nK^r = 6k, \quad k_{mnrs}K^mK^nK^rK^s = 24k.
\end{equation}
Substituting this into \eqref{psi-mass-k}, we observe that such a correction does not affect the tachyon mass at linear order in $k$. The correction is thus of the extended no-scale type, which was discussed in \cite{Berg:2007wt, Cicoli:2007xp} in the context of $\mathcal{O}(g_s^2\alpha^{\prime 2})$ corrections to type IIB Calabi-Yau orientifolds.

\subsection{Non-zero $W_{amn}$}
\label{evade3}

Let us now discuss the possibility of a superpotential with mixed couplings between the $\Phi^m$ and the $\Phi^a$ such that $W_{amn} \neq 0$. As we will see momentarily, a tachyon can then be evaded if these terms contribute at linear order in the $\lambda$-expansion,
\begin{equation}
W_{amn}^{(1)}\neq 0.
\end{equation}
In order to evade a tachyon already at order $\mathcal{O}(\lambda)$, we furthermore require
\begin{equation}
K^mK^nK^r \frac{W_{mnr}}{W}=\mathcal{O}(\lambda^2), \qquad K^m K^n \frac{W_{mn}}{W}=\mathcal{O}(\lambda^2).
\end{equation}
We can then again compute the mass of the field $\text{Re}(\e^{-i\varphi/2}\Psi) = \cos(\varphi/2) \text{Re}\Psi + \sin(\varphi/2) \text{Im}\Psi$, which we already discussed in section \ref{case3}. Under the above conditions, this becomes
\begin{align}
m^2 &= -\frac{2}{3}\left(1+\cos\varphi\right)V + \frac{1}{6} \e^G \left( \e^{-i\varphi} K_mK_nK_r \frac{\bar W^{mnr}}{\bar W} + \e^{i\varphi} K^mK^nK^r \frac{W_{mnr}}{W} \right) \nl
+ \frac{1}{6} \e^G \left( \e^{-i\varphi} (G_a+2Y_a)K_mK_n \frac{\bar W^{amn}}{\bar W} + \e^{i\varphi} (G^a+2\bar Y^a)K^mK^n \frac{W_{amn}}{W} \right) \nl +  \frac{1}{6}\e^{G}\left( \e^{-i\varphi} \frac{\bar W^{mnr}K_m K_nW_r}{W\bar W} + \e^{i\varphi} \frac{W_{mnr}K^mK^n\bar W^r}{W\bar W}\right) \nl
- \frac{1}{3}\e^{G}\left( \e^{-i\varphi} \frac{\bar W^{mn}K_m W_n}{W\bar W} + \e^{i\varphi} \frac{W_{mn}K^m\bar W^n}{W\bar W} \right) \nl 
+ \frac{1}{3} \e^{G} \left| G_a + Y_a - \e^{i\varphi} G_{ab} \left(G^b - \bar Y^b\right) \right|^2 + \mathcal{O}(\lambda^3). \label{m^2}
\end{align}
Note that this is the same expression as in \eqref{m2-case3}, except for the term due to $W_{amn}$ in the second line, which now contributes at order $\mathcal{O}(\lambda^2)$ as it is contracted with $G_a$ or $Y_a$.

The last term on the right-hand side of \eqref{m^2} vanishes if we choose $Y_a$ appropriately. The solution for $Y_a$ is then $\varphi$-dependent and takes the general form
\begin{equation}
Y_a = y_a + \e^{i\varphi}z_a, \label{ya}
\end{equation}
where $y_a, z_a$ do not depend on $\varphi$ and satisfy
\begin{equation}
G_a + y_a + G_{ab} \bar z^b = \mathcal{O}(\lambda^2), \qquad z_a - G_{ab} \left(G^b - \bar y^b\right) = \mathcal{O}(\lambda^2).
\end{equation}
Furthermore, we can write
\begin{equation}
K^mK^n\left(K^r+ \frac{\bar W^r}{\bar W}\right)\frac{W_{mnr}}{W} + (G^a+2\bar y^a)K^mK^n \frac{W_{amn}}{W} - 2 \frac{W_{mn}K^m\bar W^n}{W\bar W} = |w| \e^{i\delta}.
\end{equation}
Substituting this together with \eqref{ya} into \eqref{m^2}, we find
\begin{align}
m^2 &= -\frac{2}{3}\left(1+\cos\varphi\right)V + \frac{1}{3} \e^G |w| \cos (\varphi+\delta) + \frac{1}{3} \e^G \left( z_aK_mK_n \frac{\bar W^{amn}}{\bar W} + \bar z^aK^mK^n \frac{W_{amn}}{W} \right) \nl + \mathcal{O}(\lambda^3).
\end{align}
The first term on the right-hand side is proportional to $V$ and thus negative for all $0\le \varphi<\pi$ at a dS solution. The second term is either identically zero (for $|w|=0$) or necessarily negative for a certain $\varphi$. However, unlike in section \ref{case3}, we now also have a third term, which is due to the presence of $W_{amn}$ and does not depend on $\varphi$. Hence, one may in principle evade the tachyon, provided that the size and the sign of $W_{amn}$ can be chosen in a given model such that the third term overcompensates the negative contributions of the first two terms. As $W_{amn}$ does not enter the equations of motion, this can in general be done by dialing suitable coefficients in the superpotential.

\subsection{Unstabilized $\Phi^a$ moduli}
\label{evade1}

Another way to circumvent the appearance of the universal tachyon is to relax the assumption that the $\Phi^a$ moduli are all stabilized at the Minkowski point.
Any unstabilized modulus receives corrections to its mass at linear or higher order in the $\lambda$-expansion and is therefore in danger of becoming tachyonic at the dS solution (see, e.g., \cite{Achucarro:2015kja} for a recent analysis). One therefore expects that meta-stable dS vacua are more likely to obtain if we stabilize as many of the $\Phi^a$ as possible already at the Minkowski point. Hence, the minimal way to evade our no-go is when exactly one of the $\Phi^a$ is not stabilized at leading order. According to our discussion in section \ref{setup}, the matrix
\begin{equation}
\delta_a^c - G_a\vphantom{}^{\bar b}G_{\bar b}\vphantom{}^c \label{matrix}
\end{equation}
then has exactly one zero eigenvalue at leading order in the $\lambda$-expansion.
There are indeed examples of models where it is possible to stabilize all but one of the $\Phi^a$ at leading order \cite{Marsh:2014nla} such that there is no general obstruction to such a stabilization scheme.

It follows from our discussion of case 2 and 3 in section \ref{tachyon} that we also need to impose
\begin{equation}
K^mK^nK^r \frac{W_{mnr}}{W}=\mathcal{O}(\lambda^2), \qquad K^m K^n \frac{W_{mn}}{W}=\mathcal{O}(\lambda^2) \label{zzz}
\end{equation}
since otherwise there is a tachyon independently of whether the $\Phi^a$ are stabilized or not. As we will now explain, we then still require one additional condition in order to avoid a tachyon.

Let us go to a basis where $G_a\vphantom{}^{\bar b}$ is diagonal in the no-scale Minkowski vacuum at $\lambda=0$ and call the unstabilized direction the 1 direction. Recall that this means that one linear combination of the two real degrees of freedom in the complex $\Phi^1$ is unstabilized. Eq. \eqref{matrix} then implies that $G_1\vphantom{}^{\bar 1}G_{\bar 1}\vphantom{}^1 = 1$ and, hence, $G_1\vphantom{}^{\bar 1} = \e^{i\chi}$ at leading order in the $\lambda$-expansion, where $\chi$ is an arbitrary angle.

We now consider again the mass of a specific combination of $\text{Re}\Psi$ and $\text{Im}\Psi$, i.e., the field $\text{Re}(\e^{-i\varphi/2}\Psi) = \cos(\varphi/2) \text{Re}\Psi + \sin(\varphi/2) \text{Im}\Psi$. The mass of this field can be computed analogously to our computation in section \ref{tachyon}. This yields
\begin{align}
m^2 &= -\frac{2}{3}\left(1+\cos\varphi\right)V + \frac{1}{6} \e^G \left( \e^{-i\varphi} K_mK_nK_r \frac{\bar W^{mnr}}{\bar W} + \e^{i\varphi} K^mK^nK^r \frac{W_{mnr}}{W} \right) \nl
+  \frac{1}{6}\e^{G}\left( \e^{-i\varphi} \frac{\bar W^{mnr}K_m K_nW_r}{W\bar W} + \e^{i\varphi} \frac{W_{mnr}K^mK^n\bar W^r}{W\bar W}\right) \nl
- \frac{1}{3}\e^{G}\left( \e^{-i\varphi} \frac{\bar W^{mn}K_m W_n}{W\bar W} + \e^{i\varphi} \frac{W_{mn}K^m\bar W^n}{W\bar W} \right) \nl
+ \frac{1}{3} \e^{G} \left| G_a + Y_a - \e^{i\varphi} G_a\vphantom{}^{\bar b} \left(G_{\bar b} - \bar Y_{\bar b}\right) \right|^2 + \mathcal{O}(\lambda^3). \label{zz}
\end{align}
Note that this is the same expression as in \eqref{m2-case3}.

Now recall that a system of linear equations can always be solved as long as the eigenvalues of the coefficient matrix are non-zero. For all $a\ge 2$, we can therefore always choose a $Y_a$ such that the equation in the square is solved, again analogous to our argument in section \ref{tachyon}. Hence,
\begin{align}
m^2 &= -\frac{2}{3}\left(1+\cos\varphi\right)V + \frac{1}{6} \e^G \left( \e^{-i\varphi} K_mK_nK_r \frac{\bar W^{mnr}}{\bar W} + \e^{i\varphi} K^mK^nK^r \frac{W_{mnr}}{W} \right) \nl
+  \frac{1}{6}\e^{G}\left( \e^{-i\varphi} \frac{\bar W^{mnr}K_m K_nW_r}{W\bar W} + \e^{i\varphi} \frac{W_{mnr}K^mK^n\bar W^r}{W\bar W}\right) \nl
- \frac{1}{3}\e^{G}\left( \e^{-i\varphi} \frac{\bar W^{mn}K_m W_n}{W\bar W} + \e^{i\varphi} \frac{W_{mn}K^m\bar W^n}{W\bar W} \right) \nl
+ \frac{1}{3} \e^{G} \left| G_1 + Y_1 - \e^{i\varphi} G_1\vphantom{}^{\bar 1} \left(G_{\bar 1} - \bar Y_{\bar 1}\right) \right|^2 + \mathcal{O}(\lambda^3).
\end{align}
Let us now set $Y_1=0$ and use the above expression for $G_1\vphantom{}^{\bar 1}$ to simplify the mass term. Furthermore, we can write $G_1 = |G_1| \e^{i\gamma}$ and
\begin{equation}
K^mK^n\left(K^r+ \frac{\bar W^r}{\bar W}\right)\frac{W_{mnr}}{W} - 2 \frac{W_{mn}K^m\bar W^n}{W\bar W} = |w| \e^{i\delta}. \label{fgsigshil}
\end{equation}
We thus find
\begin{align}
m^2 &= -\frac{2}{3}\left(1+\cos\varphi\right)V + \frac{1}{3} \e^G |w| \cos (\varphi+\delta)
+ \frac{1}{3} \e^{G} \left|\left(1 - \e^{i(\varphi+\chi-2\gamma)}\right) G_1 \right|^2 + \mathcal{O}(\lambda^3). \label{fgsfg}
\end{align}
The first term on the right-hand side is proportional to $V$ and thus negative for all $0\le \varphi<\pi$ at a dS solution. In order that no linear combination of $\text{Re}\Psi$ and $\text{Im}\Psi$ is a tachyon, the sum of the other two terms thus needs to contribute positively for all these values of $\varphi$.

However, we observe that the second term, i.e., the contribution of $W_{mnr}$ and/or $W_{mn}$ and their complex conjugates, is positive only for $-\frac{\pi}{2} <\varphi+\delta< \frac{\pi}{2}$. This covers only half of the possible $\varphi$ range and can thus never suffice to remove the tachyon for all $\varphi$. We thus in any case need the third term, consistently with our discussion of case 3 in section \ref{tachyon}. On the other hand, we observe that this term vanishes for the choice $\varphi=2\gamma-\chi$. The second term needs to be positive for this choice of $\varphi$, which implies $\delta+2\gamma-\chi\in(-\frac{\pi}{2},\frac{\pi}{2})$. We conclude that there are two different cases to consider:
\begin{itemize}
 \item In the generic case $2\gamma-\chi \neq \pi$, we require two contributions to avoid a tachyon for all $\varphi$: the one from the unstabilized $\Phi^1$ and the one due to $W_{mnr}$ and/or $W_{mn}$. The superpotential thus needs to be at least quadratic in the no-scale moduli. Furthermore, we require $\delta+2\gamma-\chi\in(-\frac{\pi}{2},\frac{\pi}{2})$.
 \item For $2\gamma-\chi = \pi$, the contribution from $W_{mnr}$ and/or $W_{mn}$ is not necessary. The square in \eqref{fgsfg} then only vanishes for $\varphi=\pi$ such that it may help to stabilize the mass for all $\varphi\neq \pi$ and we cannot conclude the existence of a tachyon. The mode with $\varphi=\pi$ is then massless at quadratic order in the $\lambda$-expansion and needs to be stabilized at a higher order.
\end{itemize}
For convenience, we have summarized the different conditions under which a tachyon may be evaded in table \ref{table}.

\begin{table}[t]\renewcommand{\arraystretch}{0.9}\setlength{\tabcolsep}{4pt}
\begin{center}

\begin{tabular}{|l|l|l|l|l|l|}
\hline
&$\scriptstyle u=\mathcal{O}(\lambda^2)$ & $\scriptstyle u=\mathcal{O}(\lambda^2)$ & $\scriptstyle u=\mathcal{O}(\lambda^2)$ &	$\scriptstyle u=\mathcal{O}(\lambda^1)$& $ \scriptstyle u=\mathcal{O}(\lambda^1)$\\
& $\scriptstyle v=\mathcal{O}(\lambda^2)$&  $\scriptstyle v=\mathcal{O}(\lambda^2)$& $\scriptstyle v=\mathcal{O}(\lambda^1)$ &$\scriptstyle v=\mathcal{O}(\lambda^2)$ & $\scriptstyle v=\mathcal{O}(\lambda^1)$\\
&$\scriptstyle |w|=0$&  $\scriptstyle |w|\neq 0$ && &\\\hline
\scriptsize All $\Phi^a$ stabilized & \scriptsize Tachyon $\mathcal{O}(\lambda^2)$& \scriptsize Tachyon $\mathcal{O}(\lambda^2)$&\scriptsize Tachyon $\mathcal{O}(\lambda^1)$&\scriptsize Tachyon $\mathcal{O}(\lambda^1)$	&\scriptsize Tachyon $\mathcal{O}(\lambda^1)$\\
\scriptsize $(\det(\delta_{a}{}^{c}-G_{ab}G^{bc})\neq 0)$&&&&&\\\hline
\scriptsize One $\Phi^a$ unstabilized & \scriptsize Tachyon  $\mathcal{O}(\lambda^2)$& \scriptsize Possibly no &\scriptsize Tachyon  $\mathcal{O}(\lambda^1)$& \scriptsize            Tachyon  $\mathcal{O}(\lambda^1)$ & \scriptsize Tachyon  $\mathcal{O}(\lambda^1)$\\
\scriptsize $2\gamma-\chi\neq \pi$& &\scriptsize tachyon&&&\\
\scriptsize $(\delta+2\gamma-\chi)\in(-\frac{\pi}{2},\frac{\pi}{2})$&&&&&\\\hline
\scriptsize One $\Phi^a$ unstabilized&\scriptsize Tachyon $\mathcal{O}(\lambda^2)$&\scriptsize Tachyon $\mathcal{O}(\lambda^2)$&\scriptsize Tachyon $\mathcal{O}(\lambda^1)$&\scriptsize Tachyon $\mathcal{O}(\lambda^1)$&\scriptsize Tachyon $\mathcal{O}(\lambda^1)$\\
\scriptsize $2\gamma-\chi\neq\pi$&&&&&\\
\scriptsize $(\delta+2\gamma-\chi)\notin(-\frac{\pi}{2},\frac{\pi}{2})$&&&&&\\\hline
\scriptsize One $\Phi^a$ unstabilized&\scriptsize Possibly no ta-&\scriptsize Possibly no &\scriptsize Tachyon $\mathcal{O}(\lambda^1)$&\scriptsize Tachyon $\mathcal{O}(\lambda^1)$&\scriptsize Tachyon $\mathcal{O}(\lambda^1)$\\
\scriptsize $2\gamma-\chi=\pi$&\scriptsize chyon, but $\mathcal{O}(\lambda^2)$ &\scriptsize tachyon if &&&\\
&\scriptsize massless mode&\scriptsize $(\delta+\pi)\in(-\frac{\pi}{2},\frac{\pi}{2})$&&&
\\\hline
\scriptsize $\geq 2$ $\Phi^{a}$ unstabilized &\scriptsize Possibly no&\scriptsize Possibly no&\scriptsize Tachyon $\mathcal{O}(\lambda^1)$&\scriptsize Tachyon $\mathcal{O}(\lambda^1)$&\scriptsize Tachyon $\mathcal{O}(\lambda^{1})$\\
\scriptsize $2\gamma_A-\chi_A$ not all equal&\scriptsize tachyon&\scriptsize tachyon&&&\\\hline
\scriptsize $\geq 2$ $\Phi^a$ unstabilized&\multicolumn{5}{|c|}{\scriptsize As for one unstabilized $\Phi^a$}\\
\scriptsize $2\gamma_A-\chi_A$ all equal&\multicolumn{5}{|c|}{ }\\\hline
\end{tabular}
\caption{Stability constraints for uncorrected K{\"{a}}hler potential and $W_{amn}^{(1)}=0$, with $u=K^m K^n \frac{W_{mn}}{W}$, $v=K^mK^nK^r \frac{W_{mnr}}{W}$, $\e^{i\chi} = G_1{}^{\bar 1}$, $\e^{i\gamma}=\frac{G_1}{|G_1|}$ and $w$, $\delta$ as in \eqref{fgsigshil}.}
\label{table}
\end{center}
\end{table}

Note that, if \eqref{matrix} has more than one zero eigenvalue, the contributions from $W_{mnr}$ or $W_{mn}$ to the mass term are generically not necessary to stabilize $\Psi$. Let us count the zero eigenvalues by an index $A$. We then have $G_A\vphantom{}^{\bar A} = \e^{i\chi_A}$, $G_A = |G_A| \e^{i\gamma_A}$. Setting $W_{mnr}=W_{mn}=0$, the mass term is then given by
\begin{equation}
m^2 = -\frac{2}{3}\left(1+\cos\varphi\right)V + \frac{1}{3} \e^{G} \left| \left(1 - \e^{i(\varphi+\chi_A-2\gamma_A)}\right) G_A \right|^2 + \mathcal{O}(\lambda^3),
\end{equation}
where $\left| \ldots \right|^2$ denotes an implicit contraction with $K^{A\bar B}$. Unless $\chi_A-2\gamma_A$ takes the same value for all $A$, there is no way to choose $\varphi$ such that the square vanishes. It may therefore be possible to lift the tachyon even in models in which the superpotential is only linear in the no-scale moduli. We are not aware of a model in which this scenario is realized, but it would be interesting to explore this further.

As a notable special case, let us finally discuss models as in \cite{Marsh:2014nla}, where, in addition to \eqref{zzz}, also $W_m^{(1)}=W_{mn}^{(1)}=W_{am}^{(1)}=W_{mnr}^{(1)}=0$ holds. The equation of motion $V_a=0$ then yields $G_a+G_{ab}G^b=\mathcal{O}(\lambda^2)$ and, hence, $G_a-G_{ab}G^{bc}G_c=\mathcal{O}(\lambda^2)$. \eqref{matrix} thus necessarily has at least one zero eigenvector at leading order in the $\lambda$-expansion. In such models, our assumption that at least one $\Phi^a$ is unstabilized at the Minkowski point is therefore automatically satisfied, which confirms an observation in \cite{Marsh:2014nla}. Substituting $G_a+G_{ab}G^b=\mathcal{O}(\lambda^2)$ into \eqref{zz} together with $\varphi=\pi$ and $Y_a=0$, we furthermore find $m_{\text{Im}\Psi}^2 = -\frac{1}{6}\e^G \left(  K_mK_nK_r \frac{\bar W^{mnr}}{\bar W} + K^mK^nK^r \frac{W_{mnr}}{W} \right) + \mathcal{O}(\lambda^3)$. In the absence of a cubic term in the superpotential, such models therefore always have an unstabilized mode at quadratic order in the $\lambda$-expansion, regardless of how many zero eigenvalues \eqref{matrix} has. This is not the case in the model of \cite{Marsh:2014nla}, where a cubic term arises from a non-perturbative contribution to the superpotential.

\section{Classical dS vacua?}
\label{class-dS}

As an application of our general result, we explain in this section why classical type IIA dS vacua are difficult to obtain close to a no-scale Minkowski point. More generally, the intention of this section is to illustrate in examples how our no-go theorem may constrain or even rule out classes of string models without requiring tedious scans of the whole parameter space.

Starting with \cite{Hertzberg:2007wc, Silverstein:2007ac}, the idea of constructing simple dS vacua from classical type II flux compactifications has received a lot of interest in past years. The term ``classical dS vacua'' usually refers to models that may involve the standard NSNS, RR and metric fluxes as well as O-planes and D-branes, but no ``exotic'' ingredients such as non-geometric fluxes and no relevant perturbative or non-perturbative quantum corrections. Interestingly, all dS solutions found in such models so far are unstable \cite{Caviezel:2008tf, Flauger:2008ad, Danielsson:2009ff, Caviezel:2009tu, Danielsson:2010bc, Danielsson:2011au} even though they evade the known no-go theorems against meta-stability from the volume/dilaton moduli and the sgoldstino.

Although we do not have a general proof that classical type IIA dS vacua are impossible, our results rule them out in the important case where the $\Phi^a$ are all stabilized at the nearby no-scale Minkowski point. As we will explain below, one then requires couplings in the superpotential that are not present in classical models. The same conclusion generically applies in the case where one of the $\Phi^a$ is not stabilized at leading order. We will illustrate these points in section \ref{class-dS-2}.
Our conclusion may be evaded in non-generic cases where the superpotential of a model satisfies certain extra conditions (in particular, $2\gamma-\chi=\pi$ in section \ref{evade1}). However, we will argue in section \ref{class-dS-3} using a simple example that these conditions may be difficult to satisfy for classical models since they do not have enough tuning freedom (see also \cite{Danielsson:2012by} for similar arguments). This may explain the absence of classical dS vacua in type IIA even in such non-generic cases.

\subsection{$\le 1$ unstabilized $\Phi^a$ at leading order - generic case}
\label{class-dS-2}

Let us first discuss the case where all $\Phi^a$ are stabilized at the no-scale Minkowski point. Since corrections to the K{\"{a}}hler potential are assumed negligible in classical models, our no-go theorem can then only be evaded if $W_{amn} \neq 0$ on-shell (cf. our discussion in section \ref{evade}). This means that the superpotential needs to be at least quadratic in the no-scale moduli. A similar conclusion applies when one of the $\Phi^a$ is not stabilized at leading order but only at a higher order in the $\lambda$-expansion. According to our discussion in section \ref{evade1}, one then again generically (i.e., in the notation of section \ref{evade1}, for $2\gamma-\chi \neq \pi$) needs a superpotential which is at least quadratic in the no-scale moduli. Furthermore, the existence of a no-scale Minkowski solution requires the superpotential to be at least quadratic in the $\Phi^a$ moduli (cf. appendix \ref{app2}). A meta-stable dS solution in the vicinity of a no-scale Minkowski point thus generically requires
\begin{equation}
W_{ab} \neq 0, \qquad W_{mn} \neq 0 \label{w-conditions}
\end{equation}
off-shell.\footnote{This is true modulo a special case described at the end of appendix \ref{app2}.}
Analyzing the known expressions for flux superpotentials in type IIA string theory, one finds that the above is not possible with only NSNS, RR and metric fluxes. Although quadratic and cubic terms are generated by these ingredients, they only suffice to satisfy one of the two conditions but not both at the same time.
Stable dS vacua may therefore be allowed for superpotentials involving non-perturbative effects or non-geometric fluxes, but not for ``classical'' superpotentials in the above sense.

To see this, consider the K{\"{a}}hler potential and superpotential for a general SU(3)-structure flux compactification in type IIA (see, e.g., \cite{Grana:2005ny,Grana:2006hr,Benmachiche:2006df,Koerber:2007xk}),
\begin{align}
K &= K(z^K+ \bar z^{\bar  K}) - \ln \left[ \kappa_{ijk}(t^i + \bar t^i)(t^j + \bar t^j)(t^k + \bar t^k) \right], \\
W &= - z^K \left( i h_K + r_{iK}t^i \right) + f_6 + i f_{4i}t^i - \frac{1}{2}\kappa_{ijk}f_2^it^jt^k - \frac{i}{6}f_0 \kappa_{ijk}t^it^jt^k, \label{sp}
\end{align}
where the $t^i$ are the analogues of the K{\"{a}}hler moduli on SU(3)-structure orientifolds, the $z^K$ contain the dilaton and the analogues of the complex structure moduli, and the $\kappa_{ijk}$ are the triple-intersection numbers. The flux parameters $f_0$, $f_2^i$, $f_{4i}$, $f_6$, $h_K$ and $r_{iK}$ are related to RR, NSNS and metric fluxes, respectively.

One verifies that the only way to satisfy $W_{mn} \neq 0$ in this class of models is to take $\Phi^m=\left\{t^i\right\}$, $\Phi^a=\left\{z^K\right\}$. This implies, however, $W_{ab} = 0$. Hence, no-scale Minkowski vacua where the $t^i$ are the no-scale moduli are not possible in these compactifications. Alternatively, one may consider a subset of the $z^K$ as the no-scale moduli such that $K_mK^m=3$ is satisfied. The remaining $z^K$ and the $t^i$ are then the $\Phi^a$ moduli. In that case, $W_{ab} \neq 0$ can be satisfied such that no-scale Minkowski vacua are possible, but one then immediately sees that $W_{mn} = 0$. We therefore conclude that SU(3)-structure compactifications in type IIA cannot avoid our tachyon by turning on terms in the superpotential that are at least quadratic in the no-scale moduli. This is consistent with the absence of meta-stable dS vacua in explicit scans \cite{Caviezel:2008tf, Danielsson:2010bc, Danielsson:2011au, Danielsson:2012et, Junghans:2016uvg}.

A simple example confirming this behavior is the isotropic compactification of massive type IIA on SU(2)$\times $SU(2) which was studied in \cite{Danielsson:2010bc} (see also \cite{Caviezel:2008tf, Danielsson:2011au, Danielsson:2012et, Junghans:2016uvg}). This model admits unstable dS extrema close to a no-scale Minkowski point in moduli space at which the $\Phi^a$ moduli are all stabilized. Its K{\"{a}}hler potential and superpotential read \cite{Caviezel:2008tf, Danielsson:2012et}
\begin{align}
& K = -\ln(z_1+\bar z_1)-3 \ln(z_2+\bar z_2)-3\ln(t+\bar t)+4\ln(2), \label{geometric1} \\
& W = i \lambda_1 t^3 + 3t(\lambda_2t+z_1+z_2)-i \lambda_3 (z_1-3z_2), \label{geometric2}
\end{align}
where the $\lambda_i$ are flux numbers. Taking $\Phi^m=\left\{z_2\right\}$, $\Phi^a=\left\{t,z_1\right\}$, we observe that the superpotential is linear in the no-scale modulus.
According to our discussion of case 1 in section \ref{tachyon}, we therefore expect a tachyon along a direction in moduli space which aligns with the sgoldstino in the Minkowski limit, whereas the sgoldstino itself can be stable in the dS vacuum. This was indeed shown to be true in \cite{Junghans:2016uvg} by an explicit analysis of the mass matrix.\footnote{The Minkowski limit of this model is singular, i.e., the moduli blow up near the Minkowski point. However, this can be mapped to a regular solution by an appropriate rescaling of the fluxes.}
In \cite{Junghans:2016uvg}, the same behavior was furthermore found to be true in another example in type IIB, which also admits an unstable dS solution \cite{Caviezel:2009tu}. While we have not studied type IIB models in detail, our no-go theorem explains the appearance of the tachyons in type IIA string theory on general grounds. According to our theorem, \emph{any} classical dS extremum in type IIA close to a no-scale Minkowski point with stabilized $\Phi^a$ will have such a tachyon.

In models where the complex-structure and K{\"{a}}hler sectors of the K{\"{a}}hler potential are further separable (i.e., $K = K_1(z^A+\bar z^A)+K_2(z^\alpha+\bar z^\alpha)-\ln(t^1+\bar t^1)-\ln\left[\kappa_{1jk}(t^j+\bar t^j)(t^k+\bar t^k)\right]$ with $\left\{z^K\right\}=\left\{z^A,z^\alpha\right\}$ and $j,k=2,3,\ldots$), it is sometimes also possible to choose combinations of the $z^K$ and $t^i$ to be the no-scale moduli such that $K_mK^m=3$ is satisfied. In such models, our above arguments do therefore not immediately apply. Choosing $\Phi^m= \left\{z^A, t^k\right\}, \Phi^a = \left\{z^\alpha, t^1\right\}$, one verifies using \eqref{sp} that the equations of motion for the $\Phi^a$ moduli yield $W_{ab}=0$ at leading order such that no-scale Minkowski vacua do not exist. Choosing instead $\Phi^m= \left\{z^A, t^1\right\}, \Phi^a = \left\{z^\alpha, t^k\right\}$, no-scale Minkowski vacua are possible. However, one can check that our conclusion then still holds if all $\Phi^a$ moduli are stabilized at the Minkowski point: using again \eqref{sp}, one finds $W_{amn}=0$, which, according to our no-go theorem, implies a tachyon.\footnote{However, unlike for all other models discussed in this section, there is no simple argument addressing the case where one $\Phi^a$ modulus remains unstabilized at the Minkowski point since both $W_{ab}\neq 0$ and $W_{mn}\neq 0$ are possible in this special class of models. For the group/coset manifolds that have been analyzed in classical dS scans, the K{\"{a}}hler potential takes the simple form $K = -\ln\left[\Pi_{K=1}^4(z^K+\bar z^K)\right] -\ln\left[\Pi_{i=1}^3(t^i+\bar t^i)\right]$, and one verifies that then at least two $\Phi^a$ moduli remain unstabilized at the Minkowski point. However, for more general manifolds, this may not necessarily be the case.}

\subsection{$\ge 1$ unstabilized $\Phi^a$ at leading order - non-generic case}
\label{class-dS-3}

As discussed in section \ref{evade1}, there are (non-generic) circumstances under which it is possible to evade the tachyon even when the superpotential is only linear in the no-scale moduli, namely when one $\Phi^a$ is unstabilized with $2\gamma-\chi=\pi$ or when more than one $\Phi^a$ are unstabilized at leading order. Our no-go theorem does then not generally rule out classical dS vacua. However, satisfying the conditions for meta-stability then requires a tuning freedom that may not be available in some classical models.

Let us illustrate this in the simple model
\begin{align}
K &= -\ln(S+\bar S)-3\ln(T+\bar T)-3\ln(U+\bar U), \label{ex1} \\
W &= a_0 + ia_1U+a_2U^2+ia_3U^3 + i S \left( b_0 + ib_1U+b_2U^2+ib_3U^3 \right) \nl + i T \left( c_0 + ic_1U+c_2U^2+ic_3U^3 \right), \label{ex2}
\end{align}
which is known as the isotropic STU model. Note that, for $U=t,S=z_1,T=z_2$ and $a_0=a_1=b_2=b_3=c_2=c_3=0$, we recover the geometric type IIA example discussed around \eqref{geometric1} and \eqref{geometric2}. Some of the flux parameters $a_i,b_i,c_i$, however, correspond to non-geometric fluxes (see \cite{Danielsson:2012by} for a complete dictionary in type IIA and IIB\footnote{Our definition of the complex moduli is related to the one in \cite{Danielsson:2012by} by a factor $i$.}). Including such non-geometric fluxes, meta-stable dS vacua were found in this model in \cite{deCarlos:2009fq, deCarlos:2009qm, Danielsson:2012by, Blaback:2015zra}. In the following, we will show that this is not possible using only geometric fluxes, at least near a no-scale Minkowski point.

In order that $W_{ab}\neq 0$, we are led to consider $T$ as the no-scale modulus. We then have to set $c_i=0$ at the no-scale point such that the superpotential does not depend on $T$. The equations of motion are satisfied together with $V=0$ for
\begin{equation}
G_U = 0 = -\frac{3}{U+\bar U} + \frac{W_U}{W}, \qquad G_S = 0 = -\frac{1}{S+\bar S} + \frac{W_S}{W}.
\end{equation}
Using this in $G_{ab}$, we find the leading order expressions
\begin{align}
G_{UU} &= -  \frac{6}{(U+\bar U)^2} + \frac{W_{UU}}{W}, \quad G_{US} = -  \frac{3}{(S+\bar S)(U+\bar U)} + \frac{W_{US}}{W}, \quad G_{SS} = 0.
\end{align}
The eigenvalues of \eqref{matrix} are
\begin{align}
& 1-\frac{1}{2}G_{UU}(K^{U\bar U})^2 G_{\bar U \bar U}-G_{US}K^{S\bar S}G_{\bar S\bar U}K^{U\bar U} \nl \qquad \pm \frac{1}{2}K^{U\bar U}\sqrt{\left(G_{UU}K^{U\bar U}G_{\bar U \bar U}\right)^2 + 4G_{UU}K^{U\bar U}G_{\bar U \bar U} G_{US}K^{S\bar S}G_{\bar S\bar U}}.
\end{align}
In order to evade our no-go theorem, one or both of these eigenvalues must be zero at the Minkowski point.

Let us first consider the case where both $S,U$ are unstabilized. For finite K{\"{a}}hler metric, this yields
\begin{equation}
G_{UU} = 0, \quad G_{SU}G_{\bar S\bar U} = K_{\bar S S}K_{\bar U U}.
\end{equation}
One checks that these equations cannot be solved together with the equations of motion for the given superpotential. Hence, it is not possible to have a no-scale Minkowski vacuum with both $S,U$ unstabilized in this model.

Now we turn to the case with one unstabilized modulus. One of the two eigenvalues is zero if
\begin{equation}
\sqrt{G_{UU}G_{\bar U \bar U}} = \pm \left(G_{SU}K^{S\bar S}G_{\bar S \bar U}-K_{U\bar U}\right). \label{zero-ev}
\end{equation}
As we discussed in section \ref{evade1}, for models with one unstabilized $\Phi^a$, there is always a choice of $\varphi$ and $Y_a$ such that the equation
\begin{equation}
G_a + Y_a - \e^{i\varphi} G_{ab} \left(G^b - \bar Y^b\right) = \mathcal{O}(\lambda^2) \label{bbb}
\end{equation}
can be solved, which with $W_{mnr}=W_{amn}=W_{mn}=0$ implies a tachyon with $m^2 = -\frac{2}{3}(1+\cos\varphi)V$. The only loophole is when the equation is solved for the particular value $\varphi=\pi$. The tachyon then becomes massless at order $\lambda^2$ and might be stabilized at a higher order. Hence, we need to ask if we can solve the equation
\begin{equation}
G_a + G_{ab}G^b + Y_a - G_{ab}\bar Y^b=\mathcal{O}(\lambda^2) \label{ccc}
\end{equation}
in this model. Its components are
\begin{align}
& G_U + G_{UU}K^{U\bar U}G_{\bar U} + G_{US}K^{S\bar S}G_{\bar S} + Y_U - G_{UU}K^{U\bar U}\bar Y_{\bar U} - G_{US}K^{S\bar S}\bar Y_{\bar S}=\mathcal{O}(\lambda^2), \\
& G_S + G_{SU}K^{U\bar U}G_{\bar U} + Y_S - G_{SU}K^{U\bar U}\bar Y_{\bar U}=\mathcal{O}(\lambda^2).
\end{align}
Solving the second equation for $Y_S$ and substituting this into the first one yields
\begin{align}
& G_U + G_{UU}K^{U\bar U}G_{\bar U} + 2G_{US}K^{S\bar S}G_{\bar S} + Y_U - G_{UU}K^{U\bar U}\bar Y_{\bar U} + G_{US}K^{S\bar S}G_{\bar S\bar U}K^{U\bar U}G_{U} \nl \qquad - G_{US}K^{S\bar S}G_{\bar S\bar U}K^{U\bar U}Y_{U}=\mathcal{O}(\lambda^2).
\end{align}
Using \eqref{zero-ev} in the equation and adding to it $\mp \frac{\sqrt{G_{U U}}}{\sqrt{G_{\bar U \bar U}}}$ times its complex conjugate, we arrive at
\begin{align}
& \sqrt{G_{\bar U \bar U}}G_U \mp \sqrt{G_{U U}}G_{\bar U} + \sqrt{G_{\bar U \bar U}}G_{US}K^{S\bar S}G_{\bar S} \mp \sqrt{G_{U U}}G_{\bar U\bar S}K^{S\bar S}G_{S} =\mathcal{O}(\lambda^2), \label{ddd}
\end{align}
where the sign depends on the sign in \eqref{zero-ev}.
As expected, unlike in the case where all the moduli are stabilized, one can in general not find a $Y_a$ such that \eqref{ccc} is solved. Instead, there is one additional condition \eqref{ddd} that needs to be fulfilled (which corresponds to the condition $2\gamma-\chi=\pi$ of section \ref{evade1}).

One can check that, if only geometric fluxes are turned on, this condition cannot be satisfied together with $V>0$ in this model such that meta-stable dS vacua are forbidden, at least up to quadratic order in the $\lambda$-expansion. This statement is true both for type IIA and type IIB. The details of the computation are presented in appendix \ref{stu-model}. If, on the other hand, one adds a non-perturbative term $W_\text{np}=A\e^{-aT}$ to the superpotential, it is known that all conditions can be satisfied. In particular, it was found in \cite{Marsh:2014nla} that meta-stable dS vacua near a no-scale Minkowski point can be constructed in this model if one includes the term $W_\text{np}$. Meta-stable dS vacua near a no-scale Minkowski point were furthermore constructed in \cite{Blaback:2015zra} using non-geometric fluxes.
\\

To summarize, our no-go theorem offers an explanation for why meta-stable dS vacua of classical, geometric flux compactifications are difficult to obtain near a no-scale Minkowski point. For the case where all $\Phi^a$ are stabilized at the Minkowski point, we were in fact able to rule out such vacua in type IIA. Thus, our no-go theorem also explains the appearance of tachyons in known setups such as the type IIA example discussed around \eqref{geometric1} and \eqref{geometric2}. For the case where some of the $\Phi^a$ are unstabilized at the Minkowski point, we do not have a full proof. However, our arguments indicate that some geometric flux compactifications do not have enough free parameters to satisfy the necessary conditions to avoid our tachyon. It would be interesting to see whether this can be proven in general or whether other regions of the parameter space are less constraining. Interestingly, all of our arguments also apply to compactifications involving localized sources such as NS5-branes or KK-monopoles, which to our knowledge have not been systematically studied so far. These objects only source NSNS and metric fluxes and do therefore not lead to new terms in the superpotential.

An important caveat is that we restricted to dS vacua in the vicinity of no-scale Minkowski points in moduli space. Interesting solutions might therefore still exist far away from such points, where our systematic expansion breaks down. Furthermore, it would be interesting to see whether our results also shed light on the tachyons in SU(2)-structure compactifications in type IIB \cite{Caviezel:2009tu}, which we have not studied in detail.

\section{Conclusions}
\label{concl}

In this work, we studied string compactifications with an F-term scalar potential and analyzed to what extent they admit meta-stable dS vacua near a no-scale Minkowski point in moduli space. We showed that this is not possible for a large class of models due to a universal tachyon with second slow-roll parameter $\eta \le -\frac{4}{3}$. Our result thus also excludes slow-roll inflation for those cases. The tachyon is present unless the K{\"{a}}hler potential has sufficiently relevant $\alpha^\prime$ or $g_s$ corrections, and/or the superpotential satisfies $W_{amn}\neq 0$, and/or there is at least one unstabilized modulus at the Minkowski point perpendicular to the no-scale directions.

The direction of the tachyon in field space is model-dependent and aligns with the sgoldstino in the no-scale Minkowski limit. The sgoldstino itself, however, can be stable in the dS vacuum, as is indeed the case in explicit models. Among other applications, our result offers an explanation for why classical dS vacua in type IIA string theory are elusive, and why additional ingredients such as instanton corrections, non-geometric fluxes and/or perturbative corrections seem to be required.

Our work suggests several avenues for further research. First, it would be interesting to perform an analogous computation for models with a more general scalar potential, e.g., including constrained multiplets or D-terms. Such models were argued to admit meta-stable dS vacua, e.g., using anti-D3-branes \cite{Kachru:2003aw, Kallosh:2014wsa, Bergshoeff:2015jxa}, magnetized D7-branes \cite{Burgess:2003ic} or T-branes \cite{Cicoli:2015ylx}. It would be interesting to perform a general stability analysis of scenarios including such objects. It could also be useful to employ similar techniques to study the stability of dS solutions near other special points in moduli space which do not have a no-scale structure. Finally, it would be interesting to extend our result to all orders in $\lambda$ and thus study dS vacua and inflation far away from the no-scale Minkowski limit. We hope to come back to some of these ideas in future work.

\section*{Acknowledgements}

We would like to thank Michael Haack, Erik Plauschinn, Henry Tye, Sam Wong and Timm Wrase for useful discussions. DJ is supported by the DFG Transregional Collaborative Research Centre TRR 33 ``The Dark Universe''. The work of MZ is in part supported by the DFG Research Training Group GrK 1463 ``Analysis, Geometry and String Theory''.

\appendix

\section{No-scale Minkowski solutions in string theory}
\label{app2}

Here, we will show that no-scale Minkowski vacua in string theory necessarily have $W_{ab}\neq 0$ off-shell unless the superpotential is identically zero.
To this end, let us assume that $W_{ab} = 0$ off-shell and that the model has a no-scale Minkowski vacuum. The superpotential must then be of the form
\begin{equation}
W = c_a\Phi^a + C \label{wa1}
\end{equation}
with $c_a, C$ coefficients. Since the equations of motion yield $G_a=0$, we furthermore have
\begin{equation}
W_a = -K_a W \label{wa}
\end{equation}
at the solution. Using \eqref{wa1} in \eqref{wa}, we find
\begin{equation}
c_a = -K_a \left( c_b\Phi^b + C \right). \label{wa2}
\end{equation}

Note that non-perturbative corrections such as instantons are exponential in the moduli such that $W_{ab}\neq 0$. In order to prove our initial claim, it is therefore sufficient to focus on compactifications without such terms. The coefficients $C$ and $c_a$ are then related to fluxes in string theory. In type IIA, for example, the superpotential can receive a constant contribution from $F_6$ flux and linear terms from $H_3$ or $F_4$ fluxes. Since flux numbers are constrained to be real, $C$ and $c_a$ cannot be arbitrary complex numbers. In our conventions for the moduli, $C$ is real and the $c_a$ are imaginary (up to an overall irrelevant factor in $W$).
This can be understood from the fact that the superpotential must be invariant under a combined shift symmetry of the axions and the fluxes, which descends from a gauge symmetry of the 10d parent supergravity theory. A shift of an axion $c_a\Phi^a\to c_a(\Phi^a+i\xi_a)$ is then absorbed by a compensating shift in a flux, $C\to C-ic_a\xi^a$. This can only work if there is a relative factor $i$ between $C$ and $c_a$. Furthermore, the superpotential is constrained by dualities such as T-duality, which relate the prefactors of terms involving different types of axions.
All flux superpotentials in string theory known to us satisfy this property, and we will assume it to hold in the following.

We can then multiply \eqref{wa2} by $\Phi^a+\bar \Phi^a$ to find
\begin{equation}
c_a \left( \Phi^a+\bar \Phi^a \right) = d \left( c_b\Phi^b + C \right), \label{wa4}
\end{equation}
where we use the notation $d = - K_a (\Phi^a+\bar \Phi^a)$.
Let us first discuss the case $d=0$. We then have $c_a \left( \Phi^a+\bar \Phi^a \right)=0$. Since $c_a$ is imaginary and $C$, $K_a$ are real, it follows from \eqref{wa1} and \eqref{wa2} that $W$ is imaginary on-shell. Hence, $W=i\text{Im}\left( c_a\Phi^a + C \right)=\frac{1}{2}c_a \left( \Phi^a+\bar \Phi^a \right)=0$ on-shell. From \eqref{wa}, we then find that $W_a=c_a=0$. Substituting this back into $W$, we have $W=C=0$. Hence, the superpotential is trivial in this case.

We now discuss the general case $d\neq 0$. \eqref{wa4} then yields
\begin{equation}
C = \frac{1-d}{d} c_a\Phi^a + \frac{1}{d} c_a\bar \Phi^a. \label{wa3}
\end{equation}
Since $c_a$ is imaginary and $d$ is real, we can rewrite \eqref{wa3} into
\begin{equation}
C = -\frac{2}{d} \text{Re}\left( c_a\Phi^a \right) + \frac{2-d}{d} c_a\Phi^a. \label{wa5}
\end{equation}
For $C$ real, it then follows that $c_a\Phi^a$ is real as well and, hence, $C=-c_a\Phi^a$. From $W = c_a\Phi^a + C$ and $W_a = c_a = -K_a W$, it then follows
\begin{equation}
c_a = C = 0.
\end{equation}
Hence, we have shown that there are no no-scale Minkowski solutions in string compactifications where $W_{ab} = 0$ off-shell, except for the trivial case where the superpotential is identically (off-shell) zero.

Interestingly, one can also show that $W_{ab} \neq 0$ is a necessary condition for the $\Phi^a$ moduli to be stabilized at the no-scale Minkowski point. Using \eqref{wa}, we can write $G_{ab}=K_{ab}-K_aK_b+W_{ab}/W$ on-shell. For $W_{ab}=0$, \eqref{stab-cond} would thus reduce to
\begin{equation}
\det\left(\delta_a^c - G_{ab}G^{bc}\right) = \det\left[(2-K_bK^b)K_aK^c\right] = 0, \label{sta}
\end{equation}
where we used that $K_{ab}K^{bc}=K_{a\bar b}K^{\bar bc}=\delta_a^c$, $K_aK^{ab}=K_{\bar a}K^{\bar a b}=K^b$ and the fact that an outer product of two vectors has rank 1 and thus zero determinant. Note that the last conclusion only holds if the number of $\Phi^a$ moduli is larger than 1. However, we are not aware of any string compactification with a single $\Phi^a$ modulus. Hence, a necessary (but not sufficient) condition for the $\Phi^a$ moduli to be stabilized is that $W_{ab}$ is non-zero on-shell, which can only be true if it is also non-zero off-shell. Let us stress here that by $W_{ab} \neq 0$ we mean that $W$ is (at least) quadratic in \emph{some} of the $\Phi^a$, but not necessarily in all of them.

Finally, note that the argument below \eqref{wa5} can be avoided at a special point in moduli space where $d=-K_a (\Phi^a+\bar \Phi^a)=2$, which implies $K_aK^a=2$. However, \eqref{sta} then still applies. Furthermore, for compactification manifolds with a lot of isometries (such as group or coset manifolds), a point with $d=2$ does often not exist due to the simple form of the K{\"{a}}hler potential (see, e.g., the model discussed around \eqref{geometric1}). As another example, consider a Calabi-Yau compactification where the $\Phi^m$ are the K{\"{a}}hler moduli and the $\Phi^a$ correspond to the complex structure moduli and the dilaton. In the large complex-structure limit, one then has $-K_a (\Phi^a+\bar \Phi^a)=4$.

\section{Derivation of tachyon mass}
\label{app1}

In this appendix, we show how to compute the mass, $m^2$, of the scalar field
$\textrm{Re}(\e^{-i\varphi/2}\Psi)$, where $\Psi=T_I\Phi^I$ with $T_I=\delta_I^mK_m+\delta_I^aY_a$.
For $\varphi=0$ and $\varphi=\pi$, this gives the mass of $\textrm{Re}(\Psi)$ and $\textrm{Im}(\Psi)$, respectively.

To this end, we insert the equation of motion (\ref{eoms}) in (\ref{massmatrix1}) and (\ref{massmatrix2}) to simplify the mass matrix,
\begin{eqnarray}
V_{I\bar{J}}&=& (K_{I\bar{J}}-G_I G_{\bar{J}})V+\e^G\left[K_{I\bar{J}} +G_{IJ}G_{\bar{J}}{}^J-
G_I{}^{\bar{K}}K_{\bar{J}\bar{K}J}G^J-K_{IJ\bar{K}}G_{\bar{J}}{}^{J}G^{\bar{K}}\right.\nonumber \\
&&\left. -K_{I\bar{J}J\bar{K}}G^JG^{\bar{K}}+K_{IJ\bar{K}}K_{\bar{J}K}{}^{J}G^KG^{\bar{K}}+K_{IJ\bar{K}}K_{\bar{J}\bar{L}}{}^{\bar{K}}G^JG^{\bar{L}}
\right] , \label{Matrix1}\\
V_{IJ}&=& (G_{IJ}-G_I G_{J})V+\e^G\left[ 2G_{IJ} +G_{IJK}G^K-
G_I{}^{\bar{K}}K_{J\bar{K}K}G^K-K_{IJ\bar{K}}G^{\bar{K}}\right.\nonumber \\
&& -K_{IK\bar{K}}G_{J}{}^{\bar{K}}G^{K}-K_{IJK\bar{K}}G^KG^{\bar{K}}+K_{IL\bar{K}}K_{JK}{}^{L}G^KG^{\bar{K}}\nonumber\\
&&\left. +
K_{IK\bar{K}}K_{J\bar{L}}{}^{\bar{K}}G^KG^{\bar{L}}\right] .
\label{Matrix2}
\end{eqnarray}
As a preparation for computing
\begin{equation}
m^2=\frac{1}{2T^I T_I}	\left[2V_{\bar{I}J}T^{\bar{I}}T^J+\e^{i\varphi}V_{IJ}T^I T^J+\e^{-i\varphi}V_{\bar{I}\bar{J}}T^{\bar{I}}T^{\bar{J}}\right]\label{generalmass}
\end{equation}
from the above mass matrix, we first collect some more useful identities that follow from the equation of motion (\ref{eoms}).

\subsection{Useful identities}

As $V$ is at least of order $\lambda$, eq. (\ref{eoms}) implies for $I=a$
\begin{equation}
G_{am}K^m=-(G_a+G_{ab}G^b)+\mathcal{O}(\lambda^{2}).\label{I1}	
\end{equation}
From $K^mV_m=0$, on the other hand, one obtains
\begin{eqnarray}
&&2\left(K^m \frac{W_m}{W}+K^{\bar{m}}\frac{\bar W_{\bar m}}{\bar W}\right)+K^m K^n \frac{W_{mn}}{W}\nonumber\\
&=&-\left|\frac{W_m}{W}\right|^2-\left|K^m\frac{W_m}{W}\right|^2
-3G_aG^a-K^m\frac{W_{mn}}{W}\frac{\bar W^n}{\bar W}	\nonumber\\
&&-K^mG_{ma}G^a+\mathcal{O}(\lambda^3),\label{I2}
\end{eqnarray}
implying 
\begin{equation}
2\left(K^m \frac{W_m}{W}+K^{\bar{m}} \frac{\bar W_{\bar m}}{\bar W}\right)+K^m K^n \frac{W_{mn}}{W}=\mathcal{O}(\lambda^2).\label{I3}
\end{equation}
Contracting (\ref{I1}) with $G^a$ and eliminating $G^aG_{am}K^m$ with (\ref{I2})
gives
\begin{eqnarray}
\left|K^m \frac{W_m}{W}\right|^2&=& -2 \e^{-G}V+\left|\frac{W_m}{W}\right|^2+G^a G_{ab}G^b-K^m K^n \frac{W_{mn}}{W}\nonumber\\
&&-K^{m}\frac{W_{mn}}{W} \frac{\bar W^n}{\bar W}+\mathcal{O}(\lambda^3).	\label{I4}
\end{eqnarray}

\noindent\textbf{Special case:} $K^m K^n \frac{W_{mn}}{W}=\mathcal{O}(\lambda^2)$\\
In the special case $K^m K^n \frac{W_{mn}}{W}=\mathcal{O}(\lambda^2)$, some of the above identities simplify. 
In particular, (\ref{I3}) becomes
\begin{equation}
K^m \frac{W_m}{W}+K^{\bar{m}} \frac{\bar W_{\bar m}}{\bar W}=\mathcal{O}(\lambda^2),\label{I5}
	\end{equation}
which implies $V=\mathcal{O}(\lambda^2)$ as well as
\begin{equation}
\left(K^m \frac{W_m}{W}\right)^2=\left(K^{\bar{m}} \frac{\bar W_{\bar m}}{\bar W}\right)^2+\mathcal{O}(\lambda^3)=-\left|K^m \frac{W_m}{W}\right|^2+\mathcal{O}(\lambda^3).	\label{I6}
\end{equation}
To lowest order in $\lambda$, one furthermore obtains from $V_m=0$
\begin{equation}
\frac{W_{mn}}{W}K^n= \frac{W_m}{W} + K_{mn} \frac{\bar W^n}{\bar W}+\mathcal{O}(\lambda^2),\label{I7} 
\end{equation}
which implies
\begin{equation}
 \left|K^m\frac{W_{ml}}{W}\right|^2=\left(K^m\frac{W_{ml}}{W} \frac{\bar W^l}{\bar W}+c.c.\right)+\mathcal{O}(\lambda^3).\label{I8}
\end{equation}

\subsection{The mass}

We first consider the special case $K^mK^n\frac{W_{mn}}{W}=\mathcal{O}(\lambda^2)$, which includes the case $W_{mn}=0$ and implies via (\ref{I3}) that $V$ is at least quadratic in $\lambda$. We thus need to consider in 
general all terms in the mass (\ref{generalmass}) up to quadratic order in $\lambda$. Using (\ref{Matrix1}), one then  finds
\begin{eqnarray}
 V_{m\bar{n}}K^mK^{\bar{n}}&=& -6V + \e^G\left[ -\left(K^mK^n\frac{W_{mn}}{W}+c.c.\right)-2\left(K^m\frac{W_{ml}}{W} \frac{\bar W^l}{\bar W}+c.c. \right) \right.\nonumber\\
 &&-2\left(K^m \frac{W_m}{W}\right)^2-2\left(K^{\bar{m}} \frac{\bar W_{\bar m}}{\bar W}\right)^2-6\left|K^m \frac{W_m}{W}\right|^2 \nonumber\\
 &&\left.+\left|K^m\frac{W_{ml}}{W}\right|^2+2\left| \frac{W_m}{W}\right|^2+\left|G_a+G_{ab}G^b\right|^2 \right].
\end{eqnarray}
Using (\ref{I6}) and the real part of (\ref{I4}) to eliminate the terms in the second line and (\ref{I8}) to eliminate the first term in the third line, one ends up with
\begin{equation}
 V_{m\bar{n}}K^m K^{\bar{n}}= -2V +\e^G [G_a G^a + G_{ab}G^b G^{ac}G_{c}].\label{M1}
\end{equation}
We next determine from (\ref{Matrix2})
\begin{eqnarray}
 V_{mn}K^mK^n&=& -6V +\e^G\left[2\left(K^m \frac{W_m}{W}\right)^2-2K^m K^n\frac{W_{mn}}{W}+K^m K^n K^p\frac{W_{mnp}}{W}+2\left| \frac{W_l}{W}\right|^2\right.\nonumber\\
 &&\left. +K^m K^n \frac{W_{mnp}}{W} \frac{\bar W^p}{\bar W}+K^m K^n \frac{W_{mna}}{W}G^a-4K^m\frac{W_{mn}}{W} \frac{\bar W^n}{\bar W}\right].
 \end{eqnarray}
With (\ref{I6}) and (\ref{I4}), this becomes
\begin{eqnarray}
V_{mn}K^m K^n &=& -2V +\e^G\left[K^m K^n K^p \frac{W_{mnp}}{W}+K^m K^n \frac{W_{mnp}}{W} \frac{\bar W^p}{\bar W}\right.\nonumber \\
&&\left. +K^m K^n \frac{W_{mna}}{W}G^a-2G^aG_{ab}G^b-2K^m\frac{W_{mn}}{W} \frac{\bar W^n}{\bar W}\right]. \label{M2}
 \end{eqnarray}
Next, using (\ref{Matrix1}) and (\ref{I1}), one obtains
\begin{equation}
 V_{m\bar{a}}K^mY^{\bar{a}}=\e^G (G_{\bar{a}}+G_{\bar{a}\bar{b}}G^{\bar{b}})Y^{\bar{a}}-\e^G(G_{a}+G_{ab}G^b)Y^{\bar{a}}G_{\bar{a}}{}^{a}\label{M3}
\end{equation}
and analogously for the complex conjugate. The remaining terms that appear in $m^2$ are
\begin{eqnarray}
 V_{a\bar{b}}\bar{Y}^a Y^{\bar{b}}&=& \e^G \left[ Y_a\bar{Y}^a+\bar{Y}^aG_{ac}Y^{\bar{b}}G_{\bar{b}}{}^{c} \right], \label{M4}\\
 2V_{ma}K^m\bar{Y}^a&=&2\e^G K^mK^n\bar{Y}^a\frac{W_{mna}}{W}, \label{M5}\\
 V_{ab}\bar{Y}^a\bar{Y}^b&=&2 \e^G \bar{Y}^a\bar{Y}^bG_{ab}.\label{M6}
\end{eqnarray}
Inserting now (\ref{M1}) and (\ref{M2})--(\ref{M6}) into (\ref{generalmass}) gives 
\begin{align}
m^2 &= -\frac{2}{3}\left(1+\cos\varphi\right)V + \frac{1}{6} \e^G \left( \e^{-i\varphi} K_mK_nK_r \frac{\bar W^{mnr}}{\bar W} + \e^{i\varphi} K^mK^nK^r \frac{W_{mnr}}{W} \right) \nl
+ \frac{1}{6} \e^G \left( \e^{-i\varphi} (G_a+2Y_a)K_mK_n \frac{\bar W^{amn}}{\bar W} + \e^{i\varphi} (G^a+2\bar Y^a)K^mK^n \frac{W_{amn}}{W} \right) \nl +  \frac{1}{6}\e^{G}\left( \e^{-i\varphi} \frac{\bar W^{mnr}K_m K_nW_r}{W\bar W} + \e^{i\varphi} \frac{W_{mnr}K^mK^n\bar W^r}{W\bar W}\right) \nl
- \frac{1}{3}\e^{G}\left( \e^{-i\varphi} \frac{\bar W^{mn}K_m W_n}{W\bar W} + \e^{i\varphi} \frac{W_{mn}K^m\bar W^n}{W\bar W} \right) \nl 
+ \frac{1}{3} \e^{G} \left| G_a + Y_a - \e^{i\varphi} G_{ab} \left(G^b - \bar Y^b\right) \right|^2 + \mathcal{O}(\lambda^3), \label{Totalmass}
\end{align}
which is eq. (\ref{m^2}).

Note that when, as assumed here, $K^mK^n\frac{W_{mn}}{W}=\mathcal{O}(\lambda^2)$, the only term in the above expression that could possibly be of order $\mathcal{O}(\lambda^1)$ is the term 
involving $K^mK^nK^p\frac{W_{mnp}}{W}$ and its complex conjugate.

In the special case that $W$ is at most linear in the $\Phi^m$ and upon choosing $\varphi=0$, the above mass reduces to
\begin{align}
m_{\text{Re}\Psi}^2 &= -\frac{4}{3}V + \frac{1}{3} \e^{G} \left| G_a + Y_a - G_{ab} \left( G^{b} - \bar Y^{b} \right) \right|^2 + \mathcal{O}(\lambda^3), \label{othermass}
\end{align}
which is (\ref{tachyon-mass}).

We now finally consider the case $K^mK^n\frac{W_{mn}}{W}=\mathcal{O}(\lambda^1)$. In that case, (\ref{I3}) implies that $V=\mathcal{O}(\lambda^1)$, and one only has to keep terms linear in 
$\lambda$ in $m^2$. Going through the same steps as above, most terms then are of higher order and can be dropped leaving
\begin{align}
m^2 &= -\frac{2}{3}\left(1+\cos\varphi\right)V + \frac{1}{6} \e^G \left( \e^{-i\varphi} K_mK_nK_r \frac{\bar W^{mnr}}{\bar W} + \e^{i\varphi} K^mK^nK^r \frac{W_{mnr}}{W} \right) \nl + \mathcal{O}(\lambda^2), \label{L1mass}
\end{align}
which is (\ref{m2-case3l}). Eq. (\ref{x1}) instead follows if $\varphi=0$ is chosen and if $W_{mnp}=0$, or, more generally, 
if  $K^mK^nK^p\frac{W_{mnp}}{W}=\mathcal{O}(\lambda^2)$.

\section{STU model near a no-scale Minkowski point}
\label{stu-model}

To simplify the calculation, let us first set the moduli to 1 at the critical point, $U=S=T=1$. This can be done without loss of generality thanks to an invariance of the superpotential under a combination of shifts and rescalings of the flux numbers. One verifies that the superpotential is invariant under the change of variables $\left( S,T,U,a_i,b_i,c_i \right)\to \left( S^\prime,T^\prime,U^\prime,a_i^\prime,b_i^\prime,c_i^\prime \right)$ with
\begin{equation}
S = S^\prime \sigma_1 +i\sigma_2, \quad T = T^\prime \tau_1 +i\tau_2, \quad U = U^\prime \xi_1 +i\xi_2
\end{equation}
and
\begin{align}
& a_0^\prime = a_3\xi_2^3-b_0\sigma_2-c_3\xi_2^3\tau_2+c_2\xi_2^2\tau_2+a_0-b_3\xi_2^3\sigma_2+b_2\xi_2^2\sigma_2+b_1\xi_2\sigma_2-a_2\xi_2^2-c_0\tau_2\nl +c_1\xi_2\tau_2-a_1\xi_2, \notag \\
& a_1^\prime = -b_1\xi_1\sigma_2+3b_3\xi_1\xi_2^2\sigma_2-2b_2\xi_1\xi_2\sigma_2-2c_2\xi_1\xi_2\tau_2-c_1\xi_1\tau_2+2a_2\xi_1\xi_2-3a_3\xi_1\xi_2^2\nl+3c_3\xi_1\xi_2^2\tau_2+a_1\xi_1, \notag \\
& a_2^\prime = 3c_3\xi_2\xi_1^2\tau_2+a_2\xi_1^2-3a_3\xi_2\xi_1^2-c_2\xi_1^2\tau_2-b_2\xi_1^2\sigma_2+3b_3\xi_2\xi_1^2\sigma_2 \notag \\
& a_3^\prime = -c_3\xi_1^3\tau_2+a_3\xi_1^3-b_3\xi_1^3\sigma_2, \notag \\
& b_0^\prime = -b_2\xi_2^2\sigma_1+b_3\xi_2^3\sigma_1-b_1\xi_2\sigma_1+b_0\sigma_1, \quad
b_1^\prime = b_1\xi_1\sigma_1+2b_2\xi_1\xi_2\sigma_1-3b_3\xi_1\xi_2^2\sigma_1, \notag \\
& b_2^\prime = -3b_3\xi_1^2\xi_2\sigma_1+b_2\xi_1^2\sigma_1, \quad
b_3^\prime = b_3\xi_1^3\sigma_1, \notag \\
& c_0^\prime = c_3\xi_2^3\tau_1-c_2\xi_2^2\tau_1-c_1\xi_2\tau_1+c_0\tau_1, \quad
c_1^\prime = c_1\xi_1\tau_1+2c_2\xi_1\xi_2\tau_1-3c_3\xi_1\xi_2^2\tau_1, \notag \\
& c_2^\prime = -3c_3\xi_1^2\xi_2\tau_1+c_2\xi_1^2\tau_1, \quad
c_3^\prime = c_3\xi_1^3\tau_1,
\end{align}
where $\sigma_i,\tau_i,\xi_i$ are real parameters that can be chosen freely. Note that the K{\"{a}}hler potential transforms as $K \to K - \ln \sigma_1- 3\ln \tau_1- 3\ln \xi_1$ under this variable change. Derivatives of the K{\"{a}}hler potential are therefore not affected such that the equations of motion remain invariant.

At the no-scale Minkowski point, we require $c_i=0$ and $G_U=G_S=0$. For $U=S=T=1$, this yields
\begin{equation}
a_0 = -b_3, \quad a_1 = b_2, \quad a_2 = -b_1, \quad a_3 = b_0. \label{ng-model-eoms}
\end{equation}
Demanding that one of the $\Phi^a$ is unstabilized at the Minkowski point furthermore implies \eqref{zero-ev}. This holds if the flux parameters satisfy one of the following possible conditions:
\begin{gather}
\left\{ b_0 = 0, b_3 = 0 \right\}, \quad \left\{ b_0 = b_2, b_3 = 0 \right\}, \quad \left\{ b_1 = 0, b_2 = 0 \right\}, \notag \\ \left\{ b_1 = -3b_3, b_2 = 0 \right\}, \quad \left\{b_1 = \frac{b_3^2+b_0^2-b_2b_0}{b_3}\right\}, \quad \left\{b_0 = -\frac{1}{3}\frac{b_1^2+b_2^2+3b_3b_1}{b_2}\right\}. \label{ng-model-unstab}
\end{gather}
As an example, let us consider the first possibility. Together with \eqref{ng-model-eoms}, we thus have the choice of parameters
\begin{equation}
a_0 = 0, \quad a_1 = b_2, \quad a_2 = -b_1, \quad a_3 = 0, \quad b_0 = 0, \quad b_3 = 0
\end{equation}
at the no-scale Minkowski point.

We now attempt to construct a dS vacuum in the vicinity of the no-scale Minkowski vacuum. To this end, we perform an expansion in $\lambda$,
\begin{align}
& a_0 = a_{01}\lambda+a_{02}\lambda^2+\mathcal{O}(\lambda^3), && a_1 = b_{20}+a_{11}\lambda+a_{12}\lambda^2+\mathcal{O}(\lambda^3), \\ & a_2 = -b_{10}+a_{21}\lambda+a_{22}\lambda^2+\mathcal{O}(\lambda^3), && a_3 = a_{31}\lambda+a_{32}\lambda^2+\mathcal{O}(\lambda^3), \\ & b_0 = b_{01}\lambda+b_{02}\lambda^2+\mathcal{O}(\lambda^3), && b_1 = b_{10}+b_{11}\lambda+b_{12}\lambda^2+\mathcal{O}(\lambda^3), \\ & b_2 = b_{20}+b_{21}\lambda+b_{22}\lambda^2+\mathcal{O}(\lambda^3), && b_3 = b_{31}\lambda+b_{32}\lambda^2+\mathcal{O}(\lambda^3), \\ & c_0 = c_{01}\lambda+c_{02}\lambda^2+\mathcal{O}(\lambda^3), && c_1 = c_{11}\lambda+c_{12}\lambda^2+\mathcal{O}(\lambda^3), \\ & c_2 = c_{21}\lambda+c_{22}\lambda^2+\mathcal{O}(\lambda^3), && c_3 = c_{31}\lambda+c_{32}\lambda^2+\mathcal{O}(\lambda^3).
\end{align}
We would like to focus on solutions in which the superpotential is not trivial at leading order, i.e., we assume at least one of its flux parameters to be non-zero. We first consider the case where both $b_{10},b_{20}$ are non-zero. Note that the equations of motion are invariant under a rescaling of the superpotential by an overall factor. We can therefore set $b_{10}=1,b_{11}=0,b_{12}=0$ without loss of generality. Solving the equations of motion up to quadratic order in $\lambda$ then fixes most of the higher order coefficients. In order to evade a tachyon, we also need to satisfy the condition \eqref{ddd} at the dS solution, which fixes another coefficient. We thus find two possible solutions for the scalar potential up to quadratic order in $\lambda$:
\begin{align}
\text{solution 1: }\quad V &= 0+\mathcal{O}(\lambda^3), \notag \\
\text{solution 2: }\quad V &= -\frac{(b_{20}^2+1)(7b_{20}^2+1)(b_{20}^2+7)c_{31}^2}{24b_{20}^2(b_{20}^2+3)^2} \lambda^2+\mathcal{O}(\lambda^3).
\end{align}
Next, we consider the case where $b_{10}=0$ and $b_{20}\neq 0$. Rescaling the superpotential, we can set $b_{20}=1,b_{21}=0,b_{22}=0$. Solving again the equations of motion and \eqref{ddd}, we find
\begin{align}
\text{solution 3: }\quad V &= -\frac{(5b_{31}+c_{31}+5a_{01})(b_{31}-c_{31}+a_{01})}{192} \lambda^2+\mathcal{O}(\lambda^3).
\end{align}
Hence, $V$ can only be positive at this order if $c_{31}\neq0$, which corresponds to a non-geometric flux (see, e.g., table 1 in \cite{Danielsson:2012by}).
Finally, let us consider the remaining case where $b_{20}=0$ and $b_{10} \neq 0$. Setting $b_{10}=1,b_{11}=0,b_{12}=0$ and solving again the equations of motion together with \eqref{ddd}, we find
\begin{align}
\text{solution 4: }\quad V &= -\frac{(5b_{01}+c_{01}-5a_{31})(b_{01}-c_{01}-a_{31})}{192} \lambda^2+\mathcal{O}(\lambda^3).
\end{align}
In order that the non-geometric fluxes $c_{21}$, $c_{32}$ are zero in this solution, one furthermore finds that $b_{01}=a_{31}$ and $c_{01}=0$ is required such that $V$ vanishes. We conclude that, in the absence of non-geometric fluxes and non-perturbative corrections to the superpotential, the scalar potential cannot be made positive at this order in the $\lambda$-expansion. Analogous conclusions are reached if one considers one of the other possibilities in \eqref{ng-model-unstab}.

\bibliographystyle{utphys}
\bibliography{groups}

\end{document}